\documentclass[11pt]{article}

\usepackage{epstopdf}
\usepackage[caption=false]{subfig}
\usepackage[natbibapa,nodoi]{apacite}
\bibpunct[, ]{(}{)}{;}{a}{,}{,}

\usepackage{amsthm,amsmath,amsfonts}
\theoremstyle{plain}

\usepackage{booktabs}

\theoremstyle{definition}

\theoremstyle{remark}

\usepackage{xcolor}
\usepackage{csquotes}
\usepackage{algorithm}
\usepackage{algorithmic}
\usepackage{tikz}
\usetikzlibrary{graphs}
\usetikzlibrary{calc}
\tikzstyle{post}=[->,shorten >=1pt, >=stealth,thin]
\RequirePackage{filecontents}
\usepackage{pgfplots}
\usepackage{balance}

\usepackage[sanserif,full]{complexity}

\newcommand{\eat}[1]{}

\begin{document}

\title{Cooperative Solutions to Exploration Tasks Under Speed and Budget Constraints}

\author{Karishma \hspace{2cm} Shrisha Rao}
\date{}
\maketitle

\begin{abstract}

We present a multi-agent system where agents can cooperate to solve a
system of dependent tasks, with agents having the capability to
explore a solution space, make inferences, as well as query for
information under a limited budget. Re-exploration of the solution
space takes place by an agent when an older solution expires and is
thus able to adapt to dynamic changes in the environment. We
investigate the effects of task dependencies, with highly-dependent
graph $G_{40}$ (a well-known program graph that contains
  $40$ highly interlinked nodes, each representing a task) and
less-dependent graphs $G_{18}$ (a program graph that
  contains $18$ tasks with fewer links), increasing the speed of the
agents and the complexity of the problem space and the query budgets
available to agents. Specifically, we evaluate trade-offs between the
agent's speed and query budget. During the experiments, we observed
that increasing the speed of a single agent improves the system
performance to a certain point only, and increasing the number of
faster agents may not improve the system performance due to task
dependencies. Favoring faster agents during budget allocation enhances
the system performance, in line with the \enquote{Matthew effect.}  We
also observe that allocating more budget to a faster agent gives
better performance for a less-dependent system, but increasing the
number of faster agents gives a better performance for a
highly-dependent system.
\end{abstract}

\noindent{\bf Keywords}: Task exploration, cooperative agents,
resource constraints, multi-agent system, Matthew Effect

\section{Introduction}

Many applications like military concept development~\citep{ABM2002},
battlefield intelligence~\citep{battlefield2010,Selfregulated2018},
health care and medical diagnosis system~\citep{health2009}, etc.,
distribute tasks to achieve the goal(s). Tasks are distributed based
on the agent's capabilities, and not all the agents need to get tasks
of the same complexity.  In case of a complex task, an agent may seek
external help as well.

There also are fundamental tradeoffs involved between computation and
communication~\citep{Li2018}, as also seen in high-performance
computing (HPC)~\citep{Xiao2019}, where in some contexts it is better
to compute a solution locally and in others to fetch a solution stored
elsewhere. The same sort of tradeoff can also be seen in cloud
robotics~\citep{Garcia2015} and in 5G mobile
networks~\citep{Eramo2016}.

The Matthew Effect is also well known to exist in various forms in
various settings~\citep{rigney2010}.  However, until now, there has not
been any satisfactory simulation of the same in a broader context that
transcends the specific features of particular domains, though
attempts have been made to simulate it in specific settings, such as
scientific peer review~\citep{squazzoni2011} and computational social
systems~\citep{zhang2021}.

More generally, simulation is well known to be a useful technique to
understand tradeoffs and other aspects involved in resource
utilization strategies~\citep{dear2000,wilsdorf2019}, and to better
understand how ranking and selection may be made~\citep{waeber2012}.
However, there has not, until now, been a study of the issues involved
in how tradeoffs between speed and budgets may affect the choices
made.

Exploration by multiple agents has been known to particularly be
important in the context of multi-robot
exploration~\citep{burgard2005}, which continues to offer interesting
problems for research~\citep{viseras2020}.  However, even here, the
sorts of problems that are addressed in this work have not hitherto
been addressed at all.  Simulations of multi-robot systems have
likewise not dealt with them~\citep{dawson2010,choi2021}.

Network traffic flow evolution ~\citep{wang2018impact}, waste
collection management ~\citep{gruler2017supporting}, discrete event
systems of wireless networking ~\citep{tavanpour2020discrete},
efficient disaster management ~\citep{lee2022}, etc., are domains
where the impact of cooperation and collaboration is studied using
simulation.  In such applications, cooperation may be required among
communities or individuals in society, but collaboration may increase
the complexity.

In this work, we identify the fundamental problem of solving a set of
tasks cooperatively by a set of agents which can directly explore a 
solution space (to represent local computation) or can query an 
oracle (to represent bandwidth usage or offloaded computation), 
subject to a query budget.  The agents can also infer some new 
solutions in line with previously known solutions and can share 
their solutions with other agents.  Tasks have dependencies and need 
to be worked in an order specified by a program graph.  In this 
setting, we formulate and answer 
the following types of questions:
\begin{enumerate}

\item If there is a choice between agents with greater speed or more 
query budget, which should be preferred and why?

\item In a system of dissimilar agents operating at different speeds,
how should a fixed small budget be shared among them so that the 
overall system performance is the best possible?

\end{enumerate}

\section{Related Work}

A multi-agent system (MAS) contains multiple agents to solve complex
problems by subdividing them into smaller tasks. Agents act
autonomously to make wise decisions based on their intelligence and
experience~\citep{8352646}. In a MAS where each agent is assigned a
local task with requirements, an agent may require multiple agents'
collaboration with a coordination strategy, if
needed~\citep{Guo2017}. Interactions between tightly coupled MASs are
one of the effective means to gather the partially observable
information, while coordination policies among loosely coupled agents
is still a big challenge~\citep{Coordination2020}. Multi-agent
cooperative behavior can occur in a dynamic environment as
well~\citep{Dynamic2009} where a multi-agent cooperative processing
model performs cooperative work to process tasks quickly and
efficiently. MAS can control several aspects of smart grids like
management of energy, scheduling energy, reliability, the security of
the network, fault handling capability, communication between
agents~\citep{9299490}.

Many real-time complex systems contain task execution
dependencies~\citep{5628617}, data dependencies~\citep{6755290}, and
shared resources dependencies~\citep{8743172}.  Dependencies among
tasks also need to be noted in scheduling tasks on a system of
machines where the total energy consumed by the system is to be
reduced.  Different heuristic approaches exist for this, though the
task of energy-minimization is known to be
NP-hard~\citep{pagrawal2014}. Thus scheduling the task sets needs to
be aware of the dependencies~\citep{990595}. A dependency graph is one
of the optimal approaches to represent task
dependencies~\citep{8743172}. Execution dependencies arise among an
embedded program's tasks due to task priority, task precedence, and
inter-task communication~\citep{Crowdsourcing2019, 5676303}.

Task dependency exists during the multi-task allocation in various
applications like complex mobile crowds~\citep{Crowdsourcing2019} and
distributed computing~\citep{Lee2011}.  Unhandled dependency can cause
high latency, allocation errors and even bring the system into a wrong
state.  Thus a dependency-aware task scheduling approach is required
to obtain accuracy and efficiency.

The table scheduling algorithm, and scheduling based on task
replication can be used to schedule dependent tasks in distributed
systems~\citep{Qin2018}. The table scheduling algorithm is simple in
design and low in complexity whereas scheduling based on task
replication uses backtracking methods for task scheduling, due to
which the time complexity is high, and the solution space is quite
large.

The clustering scheduling algorithm usually divides the tasks into
smaller clusters and merges the cluster after completion. Existing
examples of clustering algorithms are EZ, DSC, LC, and
MD~\citep{Topcuoglu2002}. We have used the priority-based scheduling
algorithm with dependency constraints to eliminate the high time
complexity.

A task scheduling algorithm with resource attribute selection utilizes
the resource efficiently by selecting the optimal node to execute a
task~\citep{Zhao2014}.  This however does not consider the choice
between resources.

Task allocation is a crucial problem for agents' cooperation in
multi-agent systems. A distributed and self-adaptable scheduling
algorithm that can adapt to the task arrival process on itself,
considering the influence from task flows on other agents~\citep{8901062}.

Dominant Resource with Bottlenecked Fairness (DRBF) is a
multi-resource fair allocation mechanism to improve resource
utilization under well-studied fairness
constraints~\citep{Zhao2018}. We have evaluated the agent’s
performance with and without fair allocation of the resources, which
is useful when efficient system performance is required instead of
fairness.

Multi-agent MDP is a popular method for solving sequential
optimization, decision making, and learning problems in an uncertain
environment where the outcome depends on the previous
actions~\citep{971476}.  The presence of uncertainty regarding agent
states and actions can lead to performance issues. Policy iteration
(PI) and value iteration (VI) are the standard techniques to solve an
MDP within large action spaces~\citep{9303956}, where the iteration
complexities increase with the number of
controls~\citep{littman2013complexity,9482994}.

A partially observable Markov decision process (POMDP) is an agent
decision process for uncertainties in the planning
problem~\citep{8248668}.  A POMDP's policy is a mapping from the
observations (or belief states) to actions. However, the application
of POMDPs has been minimal for a long time because of the enormous
dimensionality and history~\citep{8576124}. Point-based
methods~\citep{kurniawati2008sarsop, shani2013survey} use heuristic
methods to find the search space and improve computational
efficiency~\citep{zhang2014covering}. Many point-based approximate
value iteration algorithms evaluate a value function to update the
estimated set of belief points~\citep{vlassis2004fast}. Subsequently,
the exploration proficiency remains to be improved, particularly when
managing large-scale POMDP applications.

The method presented here does not have issues related to uncertainty
regarding agent states similar to MDP, because an agent’s current task
exploration is independent of its previous exploration and does not
bring about an increase in dimension and history. Likewise, it is
advantageous in comparison with the POMDP, as it does not require any
value function evaluation on account of being independent of any set
of belief points. Thus it is feasible for larger applications with
longer run times.

\section{Methodology}

We present a model for a multi-agent system having cooperative agents
where tasks have some dependency structure among them, are assigned to
the system. We evaluate the solution exploration under speed and
budget constraints.

\subsection{System Specification}

Each task has a variable reward and a dependency list of other tasks on 
which it is dependent.  A task cannot be scheduled for exploration until 
all the tasks in its dependency list get explored first.  Thus, we form 
a subset of tasks whose dependency lists are empty (all tasks on which 
these tasks are dependent, have already completed).  Second, we 
prioritize tasks from this subset based on their associated rewards 
where a task with a higher value of reward gets greater priority.

Later the tasks are distributed among the available agents in the
system.  An agent can explore the solution space for an assigned task
and also collect inference data for future reference which are stored
in its knowledge base.  The advantage of the inference data is that if
an agent gets a task that can be performed using prior inference data,
then the solution space exploration is not required. A solution
provided by an agent is validated and a reward is given to an agent
based on the validation outcome.

Our model has also considered complex tasks that an agent is not able
to explore by itself and in this scenario, it can ask for help from an
oracle by making a query. Query utilization is limited as per the
allocated budget, which may be either shared or individual.  A budget
available to an agent being greater than the number of unaccomplished
tasks can eliminate the need for exploration.

When an agent gets a task assigned that belongs to its knowledge set,
it can accomplish the same quickly; however, over a while, the same
solution may no longer be valid, then an agent explores again and
updates its knowledge set. Thus the system is capable of adapting to
dynamic changes in the environment.

Agents are cooperative by sharing their knowledge with others and can
vary in terms of speed.  A cooperative faster agent is capable of
exploring the solution space and collecting inference data faster as
well. Shared knowledge from a faster agent in the early phase can
improve the performance of others as well.

For the experiments, we generated random mazes with random target 
locations.  An agent traverses the maze for the assigned task, which 
corresponds to the exploration of the solution space.  If an agent
fails to reach the solution in the generated maze, then it may query 
an oracle if it has a budget available.  The oracle provides a hint 
to explore the task in a maze instead of providing the exact 
solution.  After receiving a hint from an oracle for the
task, an agent explores the maze again and finds the solution.

Once an agent explores the solution, it also checks if the same target
location may contain solutions for other possible tasks as well.  If
so, it stores this information as inference data.  Thus, with each
current task solution found by exploration, an agent also collects
inference data.

Solution exploration is performed on a maximum $400\times400$ maze
size by multiple agents in parallel.  The obtained results show that,
as may be expected, over a while agent's knowledge increases and
improve performance by reducing the exploration time for a task.

We consider the dependencies between tasks by way of program graphs
$G_{40}$ and $G_{18}$~\citep{Zomaya2012}.  For each task in the
program graph, there is a target location defined in a maze.  It is
possible that multiple task solutions are available at the same maze
location.  If a task has a dependency on others as per the program
graph, an agent can only attempt the task, by exploration, inference,
or query, if the prior tasks are already completed.

We also evaluated the system exploration, by increasing the speed of a
single agent where the faster agent explores the solution and collects
inference data in less time.  The knowledge shared by the faster agent
can help other agents with their assigned tasks.  However, the
experimental results show that a faster agent improves the system
performance to a certain point only, due to task dependencies (see
Figure ~\ref{fig:variation}(a)).

For the experiments, we have considered two types of dependent
systems: a less-dependent system given by program graph $G_{18}$ (see
Figure~\ref{fig:G18}), and a highly-dependent one described by
$G_{40}$ (see Figure~\ref{fig:G40}). For a highly-dependent system, a
few faster agents do not have a significant impact on the average
system exploration time, but rather cause an increase the waiting time
(see Table~\ref{tab:varyFasterAgentG18}).

We evaluated a trade-off between the number of faster agents and query
budget for highly-dependent ($G_{40}$) and less-dependent ($G_{18}$)
systems (see Table~\ref{tab:varyNumber_Agents_Budget}).  Our results
show that:

\begin{enumerate}
\item It is better to increase the budget for a faster agent instead
  of increasing the number of faster agents for a less-dependent
  system.
\item It is better to increase the number of faster agents in the
  system instead of increasing the budget for a highly-dependent
  system.
\end{enumerate}

The experimental findings cover these points for the advantageous
utilization of faster agents vs. high budget. It is also seen that in
case of a limited total budget, favoring faster agents during budget
allocation improves the performance of the system (see
Table~\ref{tab:varyParameters}) in line with the \enquote{Matthew
  effect} where the rich get richer and the poor get
poorer~\citep{merton1968}.

\subsection{Model Specification}

We consider a standard model of $n$ agents in a system $A$ that is
required to $m$ tasks.  An agent $a_i \in A$. Tasks are formalized as
a 3-tuple $(T, R, D)$ where
\begin{itemize}
    \item $T = \{t_1, t_2, t_3, \ldots, t_m\}$ is a set of indivisible
      tasks, and
    \item $R = \{r_1, r_2, r_3, \ldots, r_m\}$ is a set of respective
      rewards, and
    \item $D = \{d_1, d_2, d_3, \ldots, d_m\}$ is a set of respective
      dependencies, where $d_i \subseteq T \setminus \{t_i\}$ is the
      set of tasks on which $t_i$ is dependent.
\end{itemize}

A task assignment $\lambda$ is a function $\lambda: A \rightarrow 2^T$
which indicates that a subset of tasks from $T$ is assigned to each
$a_i$.  We also require that $\lambda(a_i) \cap \lambda(a_j) =
\emptyset$, if $i \neq j$, so task assignments to different agents are
non-overlapping.

$\mu(a_i)$ is the set of tasks accomplished by $a_i$, with $\mu(a_i)
\subseteq \lambda(a_i)$.  If $\mu(a_i) = \lambda(a_i)$ then $a_i$ is
successful with all tasks assigned; else it leaves some undone.

$a_i$ can take help from an oracle by making a query. Each query to
the oracle deducts a constant amount from the allocated budget $B$,
which is a shared resource among the agents.  The oracle's help is
restricted based on available budget, and exhaustion of available
budget can lead to failure of solution space exploration.

There is a set $\mathcal{S}_j$ of possible solutions for task $t_j$.
An agent $a_i$ possesses a knowledge set $K(a_i)$ as key-value pairs,
where $t_j$ is a key and some specific $s_j \in \mathcal{S}_j$ is a
value.  The same holds for inference data as well, so for each
inferred solution $s_k$, some task $t_k$ is a key and the value is an
element $s_{k}$ of $\mathcal{S}_k$.  After a successful solution
exploration for $t_j$, $a_i$ adds the newly explored solution $s_j$,
and possibly inference data for the task to its knowledge set
$K(a_i)$.
\[ K(a_i) \leftarrow K(a_i) \cup \{(t_i,s_i)\} \cup \{(t_k,s_k)\} \]

An agent $a_{i}$ re-explores the solution for a task $t_{j}$ if an
available solution in $K(a_i)$ becomes invalid due to changes in the
environment. An agent $a_{i}$ shares its knowledge with all the other
agents.

\section{Cooperative Exploration Strategy}

This section presents the details of the exploration strategy.
Algorithm~\ref{alg:brain} describes the task scheduling, solution
validation, and update in a knowledge set among $n$
agents. Algorithm~\ref{alg:taskScheduling} filters out a set of
available tasks for the solution space exploration by considering
respective dependencies and
rewards. Algorithm~\ref{alg:enhancedKnowledgeGain} describes the
solution space exploration process by an agent.

In the algorithms, $n_e$ is an integer having the count of available
agents for solution space exploration which is initially equal to
$n$. The difference between $n$ and $n_e$ gives the count of agents
who are busy in solution space exploration.  $T_e$ is the subset of
$T$ containing the filtered tasks which do not have any unaccomplished
dependency. $T_e$ is used for the task scheduling. $R_e$ is a set of
rewards for $T_e$. $m_e$ is an integer giving the length of $T_e$, and
$\mathcal{I}$ is a set that contains the inference data in key-value
pairs where the inferred solution ($s_k$) is a value and task($t_k$)
is a key.

\begin{algorithm} 
\caption{Solution Space Exploration Algorithm}
\label{alg:brain}
\textbf{Input}: $T$: A set of tasks, $R$: A set of respective rewards, $D$: A set of respective dependencies, $n$: Number of Agents\\
\textbf{Output}: Knowledge Sets computed for all the agents
\begin{algorithmic}[1] 
\STATE  $n_e \gets n$
\STATE{// Get the independent set of tasks for exploration}
\STATE $T_{e} \gets \mathit{getAvailTasks(T, R, D, n_e )}$ \label{marker}
\STATE{// Assign the tasks to available agents}
\STATE $\mathit{taskAssignment(T_{e}, n_e)}$
\WHILE{true}
\STATE {// On receive event listener}
\STATE $\mathit{onSolnCheckMessage()}$
\STATE $t_{j}, s_j \gets \text{response from an agent } a_{i}$
\IF{$ \mathit{validateSoln(t_{j}, s_j)} $}
\STATE $\mathit{allocateReward(a_{i})}$
\STATE{// Remove the dependencies from the dependent task on the current one}
\STATE $\mathit{updateDependencies(t_{j})}$ 
\ENDIF
\STATE {// On receive event listener}
\STATE $\mathit{onTaskDoneMessage()}$
\STATE $t_{j}, s_j, \mathcal{I} \gets \text{response from an agent } a_{i}$
\STATE $K(a_i) \leftarrow K(a_i) \cup \{(t_i,s_i)\}$
\FOR{\textbf{each} $t_k, s_k \in \mathcal{I}$}
\STATE $K(a_i) \leftarrow K(a_i) \cup \{(t_k,s_k)\}$
\ENDFOR
\STATE $n_e \gets n_e + 1 $
\STATE \textbf{go to} \ref{marker}
\ENDWHILE
\end{algorithmic}
\end{algorithm}

\begin{algorithm}[]
\caption{Get available tasks for solution space exploration algorithm}
\label{alg:taskScheduling}
\textbf{Input}: $T$: A set of tasks, $R$: A set of respective rewards, $D$: A set of respective dependencies, $n_e$: Total number of available agents for solution space exploration\\
\textbf{Output}: $T_{e}$
\begin{algorithmic}[1] 
\STATE{// Filter the tasks and respective reward by eliminating the tasks which have dependencies}
\STATE $T_{e}, R_{e} \gets \mathit{getIndependentTasks(T,R,D)}$
\STATE $m_e \gets \mathit{length(T_{e})}$
\FOR{ $i \gets 0$ to $m_e-1$}
\FOR{$j \gets 0$ to $m_e-i-1$}
\IF {$R_{e}[j] < R_{e}[j+1]$}
\STATE Swap $R_{e}[j]$ and $R_{e}[j+1]$
\STATE Swap $T_{e}[j]$ and $T_{e}[j+1]$
\ENDIF
\ENDFOR
\ENDFOR
\STATE{// Return a set of available tasks for exploration}
\IF{$n_e > m_e$}
\STATE \textbf{return} $T_{e}$ \hspace{0.4cm} 
\ENDIF
\STATE \textbf{return} $T_{e}[0:n_e]$ 
\end{algorithmic}
\end{algorithm}

Algorithm~\ref{alg:brain} gets a total number of agents $n$, a set of
tasks $T$ with respective dependencies $D$, and reward $R$. In line 1
says that initially, all $n$ agents are available for solution
exploration.  In line 3, we get a set of tasks from $T_{e}$ that is
not dependent on any other task and having the highest reward.  In
line 5, each task $t_{j} \in T_{e}$ is assigned to an available agent
$a_{i}$.  In line 8, we wait for a response from an agent $a_{i}$ to
validate the explored solution by $a_i$.  In line 9, collect the
explored solution for a task $t_{j}$. In lines 10--14, we do the
validation for an explored solution and provide the respective reward
$r_{j}$ to an agent $a_{i}$ based on the validation outcomes.  The
dependency of task $t_{j}$ from all the dependent tasks on $t_{j}$ is
also removed. In line 16, we wait for a response from an agent $a_{i}$
to get the explored solution and inference data. In line 17, collect
the explored solution($s_j$) for a task $t_{j}$ and inference
data($\mathcal{I}$). In line 18, we update the $K(a_i)$ with the newly
explored solution($s_j$) as a value for a task $t_j$ as a key. In
lines 19--21, iterate through each entry in $\mathcal{I}$ which
contains inferred task($t_k$) and respective solution($s_k$) pairs,
and updates in $K(a_i)$. In line 22, we continue this process of
assignment and validation for the remaining tasks.

In Algorithm \ref{alg:brain}, line 3 uses the $\mathit{getAvailTasks}$
module, which is computed in
Algorithm~\ref{alg:taskScheduling}. Algorithm \ref{alg:taskScheduling}
accepts a total number of available agents for solution space
exploration $n_e$, a set of tasks $T$ with respective dependencies $D$
and rewards set $R$, and as an output will return a set of tasks to be
executed next. In algorithm \ref{alg:taskScheduling}, In line 2, it
returns the available tasks $T_{e}$ with respective rewards $R_{e}$
which does not have any dependency.  In lines 4--11, it sets all the
tasks in $T_e$ in descending order based on reward.  In lines 13--16,
it returns a task set $T_{e}$ when the count of available tasks
without any dependency is less than the total number of available
agents $n_e$ in the system. Otherwise, In line 14, it returns the top
$n_e$ number of tasks from the $T_{e}$.

\begin{algorithm}[]
\caption{Knowledge gain at agent algorithm}
\label{alg:enhancedKnowledgeGain}
\textbf{Input}: $t_j$: task to implement, $B$: Budget to ask queries from the oracle(Global variable)\\
\textbf{Output}: $t_j$, $s_j$, $\mathcal{I}$
\begin{algorithmic}[1] 
\STATE{// Initialize the variables to default value}
\STATE String $\mathit{hint} \gets null$
\STATE boolean $\mathit{isRewarded}=false$
\STATE$s_j, \mathcal{I}  \gets \mathit{exploreSoln(t_j, hint)}$ \label{budgetMarker}
\IF {$s_j$}
\STATE{// Check the reward status for the explored solution}
\STATE $\mathit{isRewarded} \gets \mathit{isRewardAllocated(t_j, s_j)}$
\IF{$\mathit{isRewarded}$}
\STATE{// Return task, explored solution, and inference data}
\STATE \textbf{return} $t_j, s_j, \mathcal{I}$
\ENDIF
\ELSIF{$B > 0$}
\STATE $B \gets B-1$
\STATE $\mathit{hint} \gets \mathit{askHelpFromOracle()}$
\STATE \textbf{go to} \ref{budgetMarker}
\ENDIF
\STATE \textbf{return} $null$
\end{algorithmic}
\end{algorithm}

Algorithm~\ref{alg:enhancedKnowledgeGain} describes the solution space
exploration by an agent ($a_i$).  In lines 2--3, we initialize the
variables with default values.  In line 4, the $\mathit{exploreSoln}$
function returns the explored solution $s_{j}$ and inference data for
a task $t_j$ based on the hint if provided by an oracle. Inference
data contains the set of inferred task($t_k$) with respective
solution($s_k$) in key-value pair. $\mathit{exploreSoln}$ function
also checks that if the number of unaccomplished tasks is less than
the allocated budget then directly takes help from the oracle instead
of solution exploration to reduce the exploration time. In line 7, an
agent $a_{i}$ checks the status of received reward by using
$\mathit{isRewardAllocated(t_j, s_j)}$
module. $\mathit{isRewardAllocated(t_j, s_j)}$ module sends the $s_j$
for the validation and returns a boolean value $true/false$ based on
the received reward as per the outcome of the validation. In lines
8--11, we returns the explored solution($s_j$) and inference
data($mathcal{I}$) for a task($t_j$). In lines 12--16, an agent
$a_{i}$ makes a query to an oracle if it has allocated budget greater
then zero and continue with exploration. In line 17, returns $null$ if
solution space exploration is failed.

\section{Experimental Results}

The cooperative solutions to exploration tasks strategy are checked
for multiple scenarios: even distribution of tasks across multiple
agents, average solution exploration time at the agent level, average
solution exploration time at a system level, even budget distribution,
uneven budget distribution, even speed allocation, variation in
agents' speeds, highly-dependent system, and less-dependent system.

\begin{figure}
\centering
\begin{tikzpicture}
    \node (17) at (3.5,0)[circle,draw, scale=0.8] {17};
    \node (13) at ( 1,1) [circle,draw,yshift=0.0in, scale=0.8] {13} (13.south east) edge [post] (17);
    \node (14) at ( 2.5,1) [circle,draw,yshift=0.0in, scale=0.8] {14} (14) edge [post] (17);
    \node (15) at ( 4.5,1) [circle,draw,yshift=0.0in, scale=0.8] {15} (15) edge [post] (17);
    \node (16) at ( 6,1) [circle,draw,yshift=0.0in, scale=0.8] {16} (16.south west) edge [post] (17);
    \node (9) at ( 0.5,2) [circle,draw,yshift=0.2in, scale=0.8] {9} (9.south) edge [post] (13) (9.south) edge [post] (14.north west)
    (9.south) edge [post] (15.north west) (9.south) edge [post] (16.north west);
    \node (10) at ( 2.5,2) [circle,draw,yshift=0.2in, scale=0.8] {10} (10.south) edge [post] (13.north) (10) edge [post] (14) 
    (10.south) edge [post] ($(15.north)-(0.2,0.0)$) (10.south) edge [post] ($(16.north)-(0.2,0.0)$);
    \node (11) at ( 4.5,2) [circle,draw,yshift=0.2in, scale=0.8] {11} (11.south) edge [post] ($(13.north)+(0.2,-0.05)$)
    (11.south) edge [post] (14.north east) (11) edge [post] (15) (11.south) edge [post] (16.north);
    \node (12) at ( 6.5,2) [circle,draw,yshift=0.2in, scale=0.8] {12} (12.south) edge [post] (13.north east)
    (12.south) edge [post] (14) (12.south) edge [post] (15) (12.south) edge [post] (16);
    \node (1) at ( 0,3) [circle,draw,yshift=0.25in, scale=0.8] {1} (1) edge [post] (9);
    \node (2) at ( 1,3) [circle,draw,yshift=0.25in, scale=0.8] {2} (2) edge [post] (9);
    \node (3) at ( 2,3) [circle,draw,yshift=0.25in, scale=0.8] {3} (3) edge [post] (10);
    \node (4) at ( 3,3) [circle,draw,yshift=0.25in, scale=0.8] {4} (4) edge [post] (10);
    \node (5) at ( 4,3) [circle,draw,yshift=0.25in, scale=0.8] {5} (5) edge [post] (11);
    \node (6) at ( 5,3) [circle,draw,yshift=0.25in, scale=0.8] {6} (6) edge [post] (11);
    \node (7) at ( 6,3) [circle,draw,yshift=0.25in, scale=0.8] {7} (7) edge [post] (12);
    \node (8) at ( 7,3) [circle,draw,yshift=0.25in, scale=0.8] {8} (8) edge [post] (12);
    \node (0) at ( 3.5,4) [circle,draw,yshift=0.35in, scale=0.8] {0}
    (0) edge [post] (1.north) (0) edge [post] (2.north) (0) edge [post] (3.north) (0) edge [post] (4.north)
    (0) edge [post] (5.north) (0) edge [post] (6.north) (0) edge [post] (7.north) (0) edge [post] (8.north);
\end{tikzpicture}
\caption{Task Dependency Graph $G_{18}$} \label{fig:G18}
\end{figure}
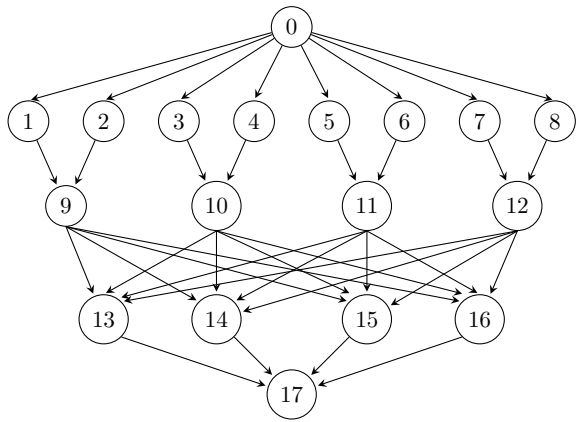

We generate a random maze with a random target location defined during
each experiment.  Maze sizes are varied.  Solution exploration is
performed on a maximum $400 \times 400$ maze size by multiple agents
in parallel.  The designed model is capable of handling the task
dependencies to simulate real-time scenarios.  We have tested the same
by using standard $G_{40}$ (Figure~\ref{fig:G40}] and $G_{18}$
  (Figure~\ref{fig:G18}) dependency program graph.  An agent $a_i$
  first explores the task on its own and may take help from the oracle
  by utilizing the allocated query budget in case of failure.  Query
  budget utilization is tested by providing the shared budget among
  the agents.  During the experiments, available tasks as per $G_{40}$
  and $G_{18}$ were distributed among 5 different cooperative agents.
  The tasks were split into multiple sets for the assignment.  In all
  results shown, exploration and waiting time unit for time is the
  second.

\begin{figure}
\centering
\begin{tikzpicture}[level distance=1in]
    \node (38) at (2,-1) [circle,draw, scale=0.8] {38};\node (39) at (5,-1) [circle,draw, scale=0.8] {39};
    \node (29) at ( 0,0) [circle,draw, scale=0.8,yshift=0.05in] {29} (29.south east) edge [post] (38);
    \node (30) at ( 1,0) [circle,draw, scale=0.8,yshift=0.05in] {30} (30.south east) edge [post] (39);
    \node (31) at ( 2,0) [circle,draw, scale=0.8,yshift=0.05in] {31} edge [post] (38);
    \node (32) at (3,0) [circle,draw, scale=0.8,yshift=0.05in] {32} (32.south east) edge [post] (39);
    \node (33) at ( 4,0) [circle,draw, scale=0.8,yshift=0.05in] {33} (33.south west) edge [post] (38);
    \node (34) at ( 5,0) [circle,draw, scale=0.8,yshift=0.05in] {34} edge [post] (39);
    \node (35) at ( 6,0) [circle,draw, scale=0.8,yshift=0.05in] {35} (35.south west) edge [post] (39);
    \node (36) at (7,0) [circle,draw, scale=0.8,yshift=0.05in] {36} (36.south west) edge [post] (39);
    \node (37) at (8,0) [circle,draw, scale=0.8,yshift=0.05in] {37} (37.south west) edge [post] (39);

    \node (20) at ( 0,1) [circle,draw, scale=0.8,yshift=0.10in] {20} edge [post] (29) edge [post] (30);
    \node (21) at ( 1,1) [circle,draw, scale=0.8,yshift=0.10in] {21} edge [post] (29) edge [post] (30) edge [post] (31);
    \node (22) at ( 2,1) [circle,draw, scale=0.8,yshift=0.10in] {22} edge [post] (30) edge [post] (31) edge [post] (32);
    \node (23) at (3,1) [circle,draw, scale=0.8,yshift=0.10in] {23} edge [post] (31) edge [post] (32) edge [post] (33);
    \node (24) at ( 4,1) [circle,draw, scale=0.8,yshift=0.10in] {24} edge [post] (32) edge [post] (33) edge [post] (34);
    \node (25) at ( 5,1) [circle,draw, scale=0.8,yshift=0.10in] {25} edge [post] (33) edge [post] (34) edge [post] (35);
    \node (26) at ( 6,1) [circle,draw, scale=0.8,yshift=0.10in] {26} edge [post] (34) edge [post] (35)edge [post] (36);
    \node (27) at (7,1) [circle,draw, scale=0.8,yshift=0.10in] {27} edge [post] (35) edge [post] (36) edge [post] (37);
    \node (28) at (8,1) [circle,draw, scale=0.8,yshift=0.10in] {28} edge [post] (36) edge [post] (37);

    \node (11) at ( 0,2) [circle,draw, scale=0.8,yshift=0.15in] {11} edge [post] (20) edge [post] (21);
    \node (12) at ( 1,2) [circle,draw, scale=0.8,yshift=0.15in] {12} edge [post] (20) edge [post] (21) edge [post] (22);
    \node (13) at ( 2,2) [circle,draw, scale=0.8,yshift=0.15in] {13} edge [post] (21) edge [post] (22) edge [post] (23);
    \node (14) at (3,2) [circle,draw, scale=0.8,yshift=0.15in] {14} edge [post] (22) edge [post] (23) edge [post] (24);
    \node (15) at ( 4,2) [circle,draw, scale=0.8,yshift=0.15in] {15} edge [post] (23) edge [post] (24) edge [post] (25);
    \node (16) at ( 5,2) [circle,draw, scale=0.8,yshift=0.15in] {16} edge [post] (24) edge [post] (25) edge [post] (26);
    \node (17) at ( 6,2) [circle,draw, scale=0.8,yshift=0.15in] {17} edge [post] (25)  edge [post] (26) edge [post] (27);
    \node (18) at (7,2) [circle,draw, scale=0.8,yshift=0.15in] {18} edge [post] (26) edge [post] (27) edge [post] (28);
    \node (19) at (8,2) [circle,draw, scale=0.8,yshift=0.15in] {19} edge [post] (27) edge [post] (28);

    \node (2) at ( 0,3) [circle,draw, scale=0.8,yshift=0.20in] {2} edge [post] (11) edge [post] (12);
    \node (3) at ( 1,3) [circle,draw, scale=0.8,yshift=0.20in] {3} edge [post] (11) edge [post] (12) edge [post] (13);
    \node (4) at ( 2,3) [circle,draw, scale=0.8,yshift=0.20in] {4} edge [post] (12) edge [post] (13) edge [post] (14);
    \node (5) at (3,3) [circle,draw, scale=0.8,yshift=0.20in] {5} edge [post] (13) edge [post] (14) edge [post] (15);
    \node (6) at ( 4,3) [circle,draw, scale=0.8,yshift=0.20in] {6} edge [post] (14) edge [post] (15) edge [post] (16);
    \node (7) at ( 5,3) [circle,draw, scale=0.8,yshift=0.20in] {7} edge [post] (15) edge [post] (16) edge [post] (17);
    \node (8) at ( 6,3) [circle,draw, scale=0.8,yshift=0.20in] {8} edge [post] (16) edge [post] (17) edge [post] (18);
    \node (9) at (7,3) [circle,draw, scale=0.8,yshift=0.20in] {9} edge [post] (17) edge [post] (18) edge [post] (19);
    \node (10) at (8,3) [circle,draw, scale=0.8,yshift=0.20in] {10} edge [post] (18) edge [post] (19);

    \node (0) at ( 2,4) [circle,draw, scale=0.8,yshift=0.25in] {0} edge [post] (2.north) edge [post] (4) edge [post] (6.north);
    \node (1) at ( 5,4) [circle,draw, scale=0.8,yshift=0.25in] {1} edge [post] (3.north) edge [post] (5.north) edge [post] (7)
    edge [post] (8.north) edge [post] (9.north) edge [post] (10.north);
\end{tikzpicture}
\caption{Task Dependency Graph $G_{40}$} \label{fig:G40}
\end{figure}
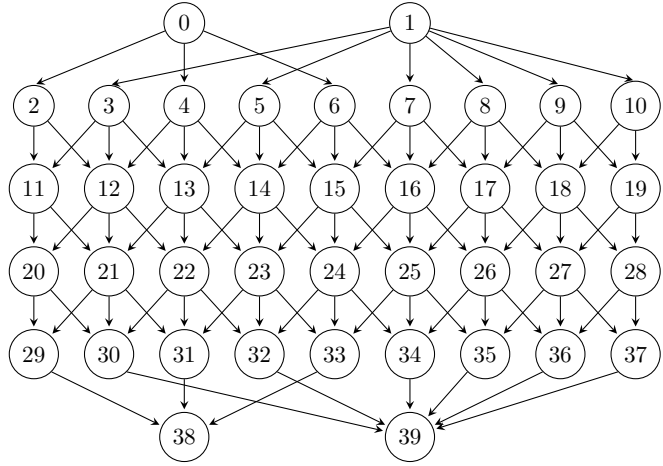

We have done multiple experiments for a $G_{40}$ program graph with 5
agents on a $400 \times 400$ maze, and observed that the average
solution exploration time taken is almost similar for all agents.  The
maze was created dynamically on each run with a random target
location.  Further, the same test is performed for more complex tasks
as well, where the agent is unable to explore the solution
independently, and takes help from an oracle, subject to a query
budget remaining.  Naturally, the average exploration time taken for a
complex task is higher in comparison to an easy task, because query
help is required by agents for complex tasks.  There is still a chance
that a task may fail even after help from an oracle, because the
oracle only provides a hint to explore the solution, instead of the
complete solution.  Not all agents in the system get complex tasks,
due to randomization and dependencies.

\begin{table}
\centering
{\begin{tabular}{cccc}
\toprule
Agents & $Expl_T(HD)$ & Total Processing Time\\
\midrule
1       & 1032.17  & 1032.17       \\
3       & 376.61  & 1129.83      \\
5       & 225.10  & 1125.5       \\
7       & 151.42  & 1059.94       \\
9       & 120.32  & 1082.88     \\
\bottomrule
\end{tabular}}
\caption{Scalability testing while varying the number of agents to
  explore the 200 tasks}
\label{tab:systemExplTimeG40}
\end{table}

In order to test the scalability of our model, we ran an experiment
varying the number of tasks keeping the number of agents constant, and
also tried another one where the number of agents is varied keeping
the number of tasks constant.  In both cases, the system graphs are of
the highly-dependent type similar to $G_{40}$.  The results clearly
indicate that for a range of values, our approach shows nearly linear
scaling.

Table~\ref{tab:systemExplTimeG40} shows the system exploration time
for a highly-dependent system like $G_{40}$ when varying the number of
agents from $1$ to $9$ for a constant $200$ tasks. And the last column
indicates that the total processing time of all the agents in the
system is consistent as the number of agents is varied, indicating
linear scaling.

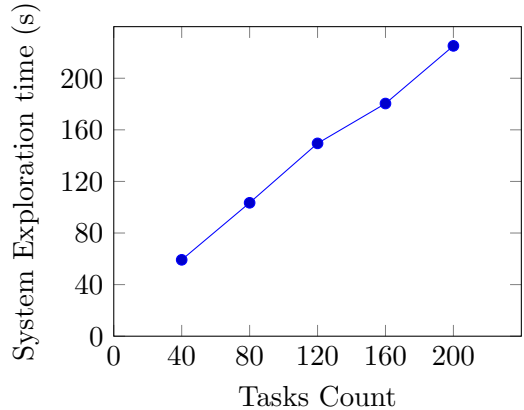
\begin{figure}
\centering
\begin{tikzpicture}
\begin{axis}[
  xlabel={Tasks Count},
  ylabel={System Exploration time (s)},
  ylabel near ticks,
  xmin=0, xmax=240,
  ymin=0, ymax=240,
  height=5.7cm,width=7cm,
  xtick={0,40,80,120, 160, 200},
  ytick={0,40,80,120, 160, 200},
  ]
  \addplot coordinates { (40, 59.21)
  (80,103.43) (120, 149.5) (160,180.4) (200,225.10)};

\end{axis}
\end{tikzpicture}
 \caption{Scalability testing while increasing the number of tasks for $5$ agents.}
 \label{fig:scalabityVaringTasksG40}
\end{figure}

Figure~\ref{fig:scalabityVaringTasksG40} shows the result of
scalability experiments when the number of tasks is varied from $40$
to $200$ for $5$ agents. It too shows linear growth for the overall
system exploration time.

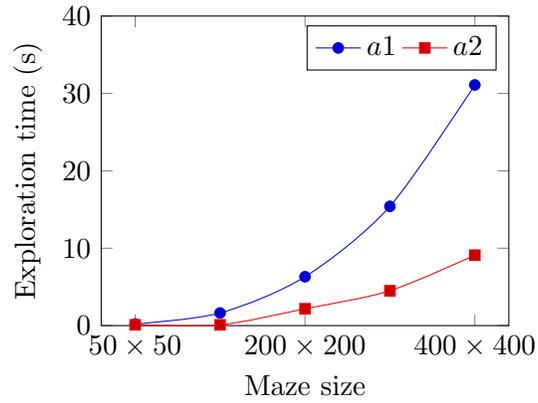
\begin{figure}
    \centering
    \pgfplotsset{table/col sep = comma}
    \begin{tikzpicture}
        \begin{axis}[
        xlabel={Maze size},
        ylabel={Exploration time (s)},
        ylabel near ticks,
        ymin=0, ymax=40,
        ytick={0,10,20,30,40},
        legend pos=north east,
        legend columns=-1
        ymajorgrids=true,
        grid style=dashed,
        height=5.7cm, width=7cm,
        xtick={0,2,...,12},
        xticklabel={
            \pgfmathparse{int(round(\tick))}
            \pgfplotstablegetelem{\pgfmathresult}{Period}\of{Data1.csv}\pgfplotsretval
        },
        ]
            \addplot+[smooth] table [x expr=\coordindex, y=spot1, col sep = comma] {Data1.csv};
            \addplot+[smooth] table [x expr=\coordindex, y=spot2, col sep = comma] {Data1.csv};
            \legend{$a1$, $a2$}
        \end{axis}
    \end{tikzpicture}
    \caption{Solution space exploration time comparison between two agents for $G_{40}$.}
 \label{fig:inferenceDataComparision}
\end{figure}

In some cases where the number of complex tasks was higher than the
allocated budget, then the number of accomplished tasks ($|\mu(i)|$)
was less than the number of assigned tasks ($|\lambda(i)|$). An agent
$a_i$ also collects inference data during explorations.  Based on
inference data, an agent's performance in terms of time execution is
(Figure~\ref{fig:inferenceDataComparision}).  \eat{Improvement in
  agent's performance is computed by comparing the exploration time of
  a new task versus a task that belongs to inference data. Time
  complexity to explore a solution in a $N \times N$ maze is
  $O(N^2)$.}



\begin{table}
\centering
{\begin{tabular}{cccc}
\toprule
Agents & $Expl_T(a_i)$ & $|\lambda(i)|$ & $TWT(i)$ \\
\midrule
1       & 30.102  & 8 & 4.17      \\
2       & 28.614  & 9 & 0      \\
3       & 29.912  & 7 & 15.01      \\
4       & 28.015  & 8 & 9.36      \\
5       & 29.721  & 9 & 4.91      \\
\bottomrule
\end{tabular}}
\caption{Waiting time due to task dependencies for $G_{40}$}
\label{tab:waitingTimeG40}
\end{table}

$Expl_T(a_i)$ stands for the average exploration time taken by an
agent $a_i$, and $TWT(a_i)$ stands for the total waiting time of an
agent $a_i$. Table~\ref{tab:waitingTimeG40} shows the total waiting
time of an agent $a_i$ due to task dependencies on other tasks by the
$G_{40}$ program graph.  With average exploration time in seconds, it
also shows the total number of assigned tasks ($|\lambda(i)|$) to an
agent $a_i$ out of $40$ tasks.  During this experiment we have
observed that $|\lambda(i)|$ was equal to $|\mu(i)|$ for all
agents. However, the higher value of the waiting time is seen to
affect $|\lambda(i)|$.  The total waiting time of individual agents
impacts the system performance.



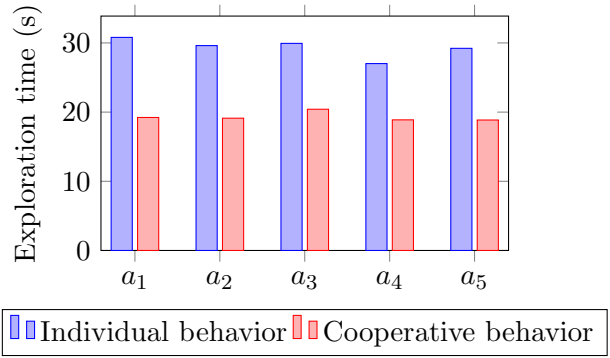
\begin{figure}
\centering
\begin{tikzpicture}
  \begin{axis}[
    ylabel=Exploration time (s),
    ylabel near ticks,
    legend style={at={(0.5,-0.25)},
    anchor=north,legend columns=-1},
    ybar, ymin=0,
    height=4.7cm, width=7cm,
    bar width=8pt,
    symbolic x coords = {$a_{1}$, $a_{2}$, $a_{3}$, $a_{4}$, $a_{5}$},
  ]
  \addplot coordinates { ($a_{1}$, 30.802) ($a_{2}$,29.614)
  ($a_{3}$,29.936) ($a_{4}$,27.015) ($a_{5}$,29.221) };
  \addplot coordinates { ($a_{1}$, 19.224) ($a_{2}$,19.122)
  ($a_{3}$,20.416) ($a_{4}$,18.880) ($a_{5}$,18.858) };
  \legend{Individual behavior, Cooperative behavior}
  \end{axis}
\end{tikzpicture}
\caption{Avg exploration time for individual vs cooperative agents for $G_{18}$}
\label{fig:barGraphg18}
\end{figure}

Figure~\ref{fig:barGraphg18} shows the average solution exploration
time taken by an agent $a_i$ for a $G_{18}$ program graph.  It shows
the average exploration time difference when agents are working
individually or cooperatively.  The experiment was performed on a $400
\times 400$ size maze, including complex tasks.  To explore the
complex tasks, the query budget is utilized by $a_i$ to take help from
an oracle.  During the experiment, the allocated budget was
insufficient to get help from an oracle for all the complex tasks.
Therefore a few complex tasks, and their dependent ones, remain
unaccomplished.

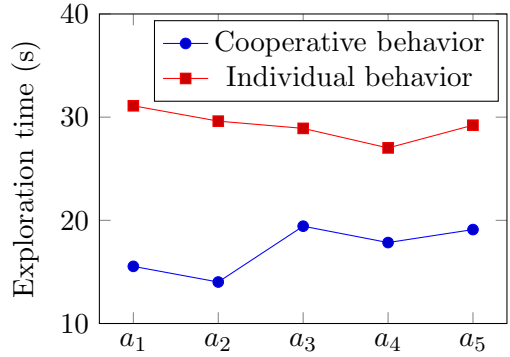
\begin{figure}
\centering
\begin{tikzpicture}
\begin{axis}[
  ylabel={Exploration time (s)},
  ylabel near ticks,
  ymin=10, ymax=40,
  height=5.7cm,width=7cm,
  ytick={0,10,20,30,40},
  symbolic x coords = {$a_{1}$, $a_{2}$, $a_{3}$, $a_{4}$, $a_{5}$},
  ]
  \addplot coordinates { ($a_{1}$, 15.539) ($a_{2}$,14.021)
  ($a_{3}$,19.433) ($a_{4}$,17.845) ($a_{5}$,19.104) };
  \addplot coordinates { ($a_{1}$, 31.102) ($a_{2}$,29.614)
  ($a_{3}$,28.912) ($a_{4}$,27.015) ($a_{5}$,29.221) };

\legend{Cooperative  behavior, Individual behavior}
\end{axis}
\end{tikzpicture}
 \caption{Cooperative faster agent improves the system performance for $G_{18}$.}
 \label{fig:fasterAgentSystemPerformanceG18}
\end{figure}

Figure \ref{fig:fasterAgentSystemPerformanceG18} shows that a faster
agent improves the performance where some tasks were related to
others' inference data and are completed due to said inference
data. Faster agents have shared the inference data with others and
reduced the exploration times for other agents as well.



Figure~\ref{fig:inferenceDataComparision} shows evaluations across
several maze sizes.  It shows two different agent behaviors out of 5
where agent $a_1$ gets new tasks and does the solution exploration,
whereas agent $a_2$ gets the tasks for which solutions are already
available due to inference data.  The inference data was either
collected by agent $a_2$ during the task exploration, or received from
other agents in the system.  Figure~\ref{fig:inferenceDataComparision}
clearly shows that an agent $a_2$ takes less time for the solution
space exploration while compared with exploration time taken by an
agent $a_1$.  $a_2$'s exploration time is approximately $70\%$ less in
comparison with $a_1$'s exploration time.


\begin{table}
\centering
{\begin{tabular}{cccc}
\toprule
$f$ & $Expl_T(LD)$ & $WT$ \\
\midrule
0 & 28.97  &  4.60     \\
1 & 28.31  &  7.20     \\
2 & 28.06  &  7.41     \\
3 & 27.65  &  7.70     \\
4 & 20.84  &  2.15     \\
5 & 19.57  &  1.42     \\
\bottomrule
\end{tabular}}
\caption{System performance when varying the count of faster agents for $G_{18}$}
\label{tab:varyFasterAgentG18}
\end{table}

Table~\ref{tab:varyFasterAgentG18} shows the observations when we vary
the number of faster agents $f$ out of 5 in the system where a faster
agent's speed was $2\times$ while comparing with others. $Expl_T(LD)$
stands for the average waiting time for a less-dependent system, and
$WT$ stands for a system's average waiting time.  We observe that
system performance improves when $f \geq 4$ for the $G_{18}$ program
dependency graph.  That shows the system performance, which is
dependent on available faster agents, varies based on the task
dependencies.  Fewer faster agents cannot improve the system
performance due to pending exploration for parent tasks from slower
agents; we just see an increase in the average waiting time of the
system due to an increase in waiting times of the faster agents.


\begin{figure}
    \centering
    \pgfplotsset{table/col sep = comma}
    \begin{tikzpicture}
        \begin{axis}[
        ylabel={Exploration time (s)},
        ylabel near ticks,
        ymin=20, ymax=32,
        ytick={15,20,25,30,35},
        height=5cm,width=4.5cm,
        xticklabel={
            \pgfmathparse{int(round(\tick))}
            \pgfplotstablegetelem{\pgfmathresult}{Period}\of{Data2.csv}\pgfplotsretval
        },
        ]
            \addplot+[smooth] table [x expr=\coordindex, y=spot1, col sep = comma] {Data2.csv};
        \end{axis}
        \label{fig:varySpeed_an_agent_fixedBudget}
        \node[below] at (1.3,-0.55) {(a) Varying Speed};
    \end{tikzpicture} %
    \begin{tikzpicture}
        \begin{axis}[
            ylabel=Exploration time (s),
            ylabel near ticks,
            height=5cm,width=4.5cm,
            ymin=10,ymax=50,
            symbolic x coords = {$1x$, $2x$, $3x$, $4x$,$5x$},
          ]
            \addplot coordinates { ($1x$,29.83 ) ($2x$,26.07) ($3x$,25.31) ($4x$,24.40) ($5x$,23.43) };
            \addplot coordinates { ($1x$, 26.54) ($2x$,23.53) ($3x$,22.34) ($4x$,21.52) ($5x$,21.03) };
            \addplot coordinates { ($1x$, 22.30) ($2x$,19.11) ($3x$,17.64) ($4x$,16.30) ($5x$,15.43) };
            \legend{Budget:20, Budget:40, Budget:80}
    \end{axis}
    \label{fig:linearImprovement}
    \node[below] at (1,-0.55) {(b) Varying Speed + Budget};
    \end{tikzpicture}
    
    \caption{Varying speed and budget of an agent in a system  $G_{40}$.}
      \label{fig:variation}
\end{figure}
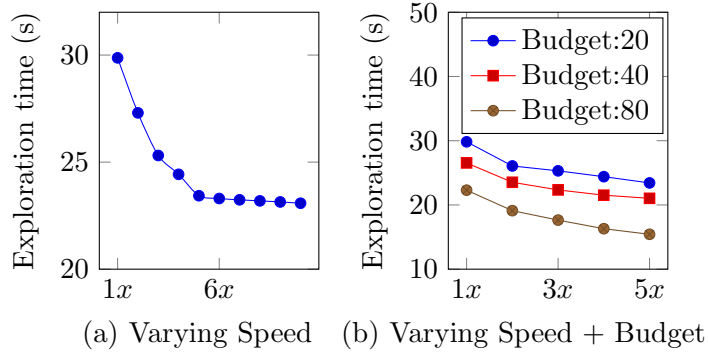

Figure~\ref{fig:variation}(a) shows that increasing the speed of an
agent $a_i$ in a highly-dependent system like $G_{40}$ initially
improves the performance, but due to dependencies, the performance
becomes constant after a specific speed increment.  Varying the budget
in increasing order for different speed agents improves the individual
agent's performance consistently, as shown in
Figure~\ref{fig:variation}(b), where we have tested the performance
with a budget of 20, 40, or 80 to 5 agents of different speeds.

\begin{table}
\centering
{\begin{tabular}{cccc}
\toprule
$f$ & Budget &$Expl_T(LD)$ & $Expl_T(HD)$\\
\midrule
0 & 20  &  29.73 &  29.87     \\
1 & 20  &  27.30 &  28.52      \\
1 & 40  &  25.95 &  28.13      \\
1 & 60  &  24.54 &  26.94      \\
\textbf{1} & \textbf{80}  &  \textbf{23.02} &  26.33    \\
\midrule
1 & 100  &  21.73 &  25.54     \\
2 & 100  &  19.41 &  22.41     \\
3 & 100  &  17.70 &  19.38     \\
4 & 100  &  14.79 &  15.86     \\
\midrule
2 & 20  &  26.48  &  26.96   \\
3 & 20  &  24.51  &  25.79     \\
\textbf{4} & \textbf{20}  &  23.90  &  \textbf{24.64}  \\
\bottomrule
\end{tabular}}
\caption{System performance when varying the number of faster agents
  and budget for a less-dependent and highly-dependent system.}
\label{tab:varyNumber_Agents_Budget}
\end{table}

We evaluated a trade-off between several faster agents vs. query
budget for a highly-dependent and less-dependent system, as shown in
Table~\ref{tab:varyNumber_Agents_Budget}.  $Expl_T(LD)$ stands for the
average exploration time for the less-dependent system $G_18$, and
$Expl_T(HD)$ stands for the average exploration time for the
highly-dependent system $G_{40}$.  We observe that a high budget (80)
for a single faster agent reduces the exploration time to $23.02$.  In
contrast, an increment in the number of faster agents reduces the
exploration time to $23.90$ for a less-dependent system.  Thus it is
better to increase the budget for a faster agent, instead of
increasing the number of faster agents, in a less-dependent
system. Similarly, an increment in the number of faster agents for a
highly-dependent system reduces the exploration time to $24.64$.  Thus
it is better to increase the number of faster agents in the system
instead of increasing the budget for a highly-dependent system.

\begin{table}
\centering  
{\begin{tabular}{cccc}
\toprule
Scenarios & $Expl_T(LD)$ & $Expl_T(HD)$ \\
\midrule
 1 & 19.81 & 21.63\\
\midrule
 2 & 15.97 & 19.75\\ 
\midrule
 3 & 13.78 & 17.66\\
\midrule
 4 & 10.92 & 14.24\\ 
\midrule
 5  & 10.07 & 13.97\\
\bottomrule
\end{tabular}}
\caption{System performance when varying the budget for dissimilar agents
  for a less-dependent and highly-dependent system}
\label{tab:varyParameters}
\end{table}

Table~\ref{tab:varyParameters} shows the exploration times for a
less-dependent and highly-dependent system where 5 different-speed
agents are present.  We also evaluate the average exploration time
while allocating dissimilar budgets to an individual agent.  Speed and
budget combinations for $Scenario1$ is ($1\times$, 45), ($2\times$,
25), ($3\times$, 15), ($4\times$, 10), ($5\times$, 5), for $Scenario2$
is ($1\times$, 30), ($2\times$, 25), ($3\times$, 20), ($4\times$, 15),
($5\times$, 10), for $Scenario3$ is ($1\times$, 20), ($2\times$, 20),
($3\times$, 20), ($4\times$, 20), ($5\times$, 20), for $Scenario4$ is
($1\times$, 10), ($2\times$, 15), ($3\times$, 20), ($4\times$, 25),
($5\times$, 30) and for $Scenario5$ is ($1\times$, 5), ($2\times$,
10), ($3\times$, 15), ($4\times$, 25), ($5\times$, 45).  The
evaluation results of all scenarios shows that the system exploration
time reduces when favoring faster agents, in line with the
\enquote{Matthew effect}~\citep{merton1968}, for both less-dependent
and highly-dependent systems.

In summary, the following are key findings of our work:

\begin{enumerate}
  
\item Agents' performance improves due to collection of inference data
  (Figure~\ref{fig:inferenceDataComparision}).  This is in line with
  prior work that shows that using inference improves performance in
  goal-oriented collaborative work~\citep{Liu2016}.
  
\item Cooperative behavior of agents improves the agents' performance
  (Figure~\ref{fig:barGraphg18}) as well as the system performance as
  a whole (Figure~\ref{fig:fasterAgentSystemPerformanceG18}).  It is
  well known that cooperation improves motivation~\citep{Carr2014},
  but our work suggests that it improves performance even when
  psychological aspects are not involved.

\item Increasing speeds of agents improves the system performance up
  to a certain point only.  Due to dependency on other tasks it may
  not improve the performance further (Figure ~\ref{fig:variation}a).
  In some cases it may increase the waiting time of an agent. Where an
  agent will wait for other task to be available for exploration
  (Table 3).  This is in line with Amdahl's Law for parallel
  processing~\citep{Hill2008} which also holds that increasing the
  speed of a single component in a multi-processor system does not
  improve system performance beyond a point.

\item Increasing speed and budget for an agent, linearly improves the
  system performance (Figure ~\ref{fig:variation}b).
  
\item Constraints evaluation for highly dependent and less dependent
  system shows that its better to increase number of faster agent in a
  highly dependent system, while it is better to increase budget in a
  less-dependent system (Table ~\ref{tab:varyNumber_Agents_Budget}).

\item Increasing budget for a faster agent gives the better system
  performance (Table ~\ref{tab:varyParameters}).  This is in line with
  the \enquote{Matthew Effect}~\citep{merton1968} that also holds that
  it is better to reward the higher-performing, rather than to spread
  resources equitably.
\end{enumerate}

\section{Conclusions}

In this paper, we have evaluated trade-offs between agents'
constraints of speed and query budget for a system where agents are
dissimilar in speed but similar in function, and can solve problems
directly as well as by querying.  As shown in our experimental
results, favoring faster agents during budget allocation with a fixed
total budget reduces the exploration time efficiently, in line with
the \enquote{Matthew effect.}  The experimental findings showed that
allocating more budget to a faster agent offers better performance in
a less-dependent system, while in a highly-dependent system,
increasing the number of faster agents offers a better performance.

Given the large number of systems where solutions to complex problems
can be computed cooperatively by several agents, or gained by query or
inference subject to constraints, weaver that this work can be used to
formulate a set of guidelines for improving the performances of such
systems given necessary trade-offs.

Currently, a static reward value is used for task prioritization. The
limitation of the system is, the reward is not reducing or expiring
over a period of time. which is not inline with the hard real-time
applications like flight control systems, nuclear power plants, stock
exchange, medical and automotive equipment~\citep{776398}. Future work
concerns the adaption of the proposed solution for hard real-time
application where time plays a critical role for the explored
solution. One objective is to consider the varying reward, which is
reducing over a period of time.


\begin{thebibliography}{}

\bibitem [\protect \citeauthoryear {%
Agrawal%
\ \BBA {} Rao%
}{%
Agrawal%
\ \BBA {} Rao%
}{%
{\protect \APACyear {2014}}%
}]{%
pagrawal2014}
\APACinsertmetastar {%
pagrawal2014}%
\begin{APACrefauthors}%
Agrawal, P.%
\BCBT {}\ \BBA {} Rao, S.%
\end{APACrefauthors}%
\unskip\
\newblock
\APACrefYearMonthDay{2014}{{\APACmonth{10}}}{}.
\newblock
{\BBOQ}\APACrefatitle {Energy-Aware Scheduling of Distributed Systems}
  {Energy-aware scheduling of distributed systems}.{\BBCQ}
\newblock
\APACjournalVolNumPages{{IEEE} Trans. Autom. Sci. Eng.}{11}{4}{1163--1175}.
\newblock
\APACrefnote{doi:10.1109/TASE.2014.2308955}
\newblock
\begin{APACrefDOI} \doi{10.1109/TASE.2014.2308955} \end{APACrefDOI}
\PrintBackRefs{\CurrentBib}

\bibitem [\protect \citeauthoryear {%
Anceaume%
, Cabillic%
, Chevochot%
\BCBL {}\ \BBA {} Puaut%
}{%
Anceaume%
\ \protect \BOthers {.}}{%
{\protect \APACyear {1999}}%
}]{%
776398}
\APACinsertmetastar {%
776398}%
\begin{APACrefauthors}%
Anceaume, E.%
, Cabillic, G.%
, Chevochot, P.%
\BCBL {}\ \BBA {} Puaut, I.%
\end{APACrefauthors}%
\unskip\
\newblock
\APACrefYearMonthDay{1999}{}{}.
\newblock
{\BBOQ}\APACrefatitle {A flexible run-time support for distributed dependable
  hard real-time applications} {A flexible run-time support for distributed
  dependable hard real-time applications}.{\BBCQ}
\newblock
\BIn{} \APACrefbtitle {Proceedings 2nd IEEE International Symposium on
  Object-Oriented Real-Time Distributed Computing (ISORC'99) (Cat.
  No.99-61702)} {Proceedings 2nd ieee international symposium on
  object-oriented real-time distributed computing (isorc'99) (cat.
  no.99-61702)}\ (\BPG~310-319).
\newblock
\begin{APACrefDOI} \doi{10.1109/ISORC.1999.776398} \end{APACrefDOI}
\PrintBackRefs{\CurrentBib}

\bibitem [\protect \citeauthoryear {%
Ashutosh%
\ \protect \BOthers {.}}{%
Ashutosh%
\ \protect \BOthers {.}}{%
{\protect \APACyear {2020}}%
}]{%
9303956}
\APACinsertmetastar {%
9303956}%
\begin{APACrefauthors}%
Ashutosh, K.%
, Consul, S.%
, Dedhia, B.%
, Khirwadkar, P.%
, Shah, S.%
\BCBL {}\ \BBA {} Kalyanakrishnan, S.%
\end{APACrefauthors}%
\unskip\
\newblock
\APACrefYearMonthDay{2020}{}{}.
\newblock
{\BBOQ}\APACrefatitle {Lower Bounds for Policy Iteration on Multi-action MDPs}
  {Lower bounds for policy iteration on multi-action mdps}.{\BBCQ}
\newblock
\BIn{} \APACrefbtitle {2020 59th IEEE Conference on Decision and Control (CDC)}
  {2020 59th ieee conference on decision and control (cdc)}\ (\BPG~1744-1749).
\newblock
\begin{APACrefDOI} \doi{10.1109/CDC42340.2020.9303956} \end{APACrefDOI}
\PrintBackRefs{\CurrentBib}

\bibitem [\protect \citeauthoryear {%
Burgard%
, Moors%
, Stachniss%
\BCBL {}\ \BBA {} Schneider%
}{%
Burgard%
\ \protect \BOthers {.}}{%
{\protect \APACyear {2005}}%
}]{%
burgard2005}
\APACinsertmetastar {%
burgard2005}%
\begin{APACrefauthors}%
Burgard, W.%
, Moors, M.%
, Stachniss, C.%
\BCBL {}\ \BBA {} Schneider, F.%
\end{APACrefauthors}%
\unskip\
\newblock
\APACrefYearMonthDay{2005}{}{}.
\newblock
{\BBOQ}\APACrefatitle {Coordinated multi-robot exploration} {Coordinated
  multi-robot exploration}.{\BBCQ}
\newblock
\APACjournalVolNumPages{IEEE Transactions on Robotics}{21}{3}{376-386}.
\newblock
\begin{APACrefDOI} \doi{10.1109/TRO.2004.839232} \end{APACrefDOI}
\PrintBackRefs{\CurrentBib}

\bibitem [\protect \citeauthoryear {%
{Cares}%
}{%
{Cares}%
}{%
{\protect \APACyear {2002}}%
}]{%
ABM2002}
\APACinsertmetastar {%
ABM2002}%
\begin{APACrefauthors}%
{Cares}, J\BPBI R.%
\end{APACrefauthors}%
\unskip\
\newblock
\APACrefYearMonthDay{2002}{}{}.
\newblock
{\BBOQ}\APACrefatitle {The use of agent-based models in military concept
  development} {The use of agent-based models in military concept
  development}.{\BBCQ}
\newblock
\BIn{} \APACrefbtitle {Proceedings of the Winter Simulation Conference}
  {Proceedings of the winter simulation conference}\ (\BVOL~1, \BPG~935-939
  vol.1).
\newblock
\begin{APACrefDOI} \doi{10.1109/WSC.2002.1172983} \end{APACrefDOI}
\PrintBackRefs{\CurrentBib}

\bibitem [\protect \citeauthoryear {%
Carr%
\ \BBA {} Walton%
}{%
Carr%
\ \BBA {} Walton%
}{%
{\protect \APACyear {2014}}%
}]{%
Carr2014}
\APACinsertmetastar {%
Carr2014}%
\begin{APACrefauthors}%
Carr, P\BPBI B.%
\BCBT {}\ \BBA {} Walton, G\BPBI M.%
\end{APACrefauthors}%
\unskip\
\newblock
\APACrefYearMonthDay{2014}{}{}.
\newblock
{\BBOQ}\APACrefatitle {Cues of working together fuel intrinsic motivation}
  {Cues of working together fuel intrinsic motivation}.{\BBCQ}
\newblock
\APACjournalVolNumPages{Journal of Experimental Social
  Psychology}{53}{}{169--184}.
\newblock
\begin{APACrefDOI} \doi{10.1016/j.jesp.2014.03.015} \end{APACrefDOI}
\PrintBackRefs{\CurrentBib}

\bibitem [\protect \citeauthoryear {%
Choi%
\ \protect \BOthers {.}}{%
Choi%
\ \protect \BOthers {.}}{%
{\protect \APACyear {2021}}%
}]{%
choi2021}
\APACinsertmetastar {%
choi2021}%
\begin{APACrefauthors}%
Choi, H.%
, Crump, C.%
, Duriez, C.%
, Elmquist, A.%
, Hager, G.%
, Han, D.%
\BDBL {}Trinkle, J.%
\end{APACrefauthors}%
\unskip\
\newblock
\APACrefYearMonthDay{2021}{}{}.
\newblock
{\BBOQ}\APACrefatitle {On the use of simulation in robotics: Opportunities,
  challenges, and suggestions for moving forward} {On the use of simulation in
  robotics: Opportunities, challenges, and suggestions for moving
  forward}.{\BBCQ}
\newblock
\APACjournalVolNumPages{Proceedings of the National Academy of
  Sciences}{118}{1}{}.
\newblock
\begin{APACrefURL} \url{https://www.pnas.org/content/118/1/e1907856118}
  \end{APACrefURL}
\newblock
\begin{APACrefDOI} \doi{10.1073/pnas.1907856118} \end{APACrefDOI}
\PrintBackRefs{\CurrentBib}

\bibitem [\protect \citeauthoryear {%
David%
, Cottet%
\BCBL {}\ \BBA {} Nissanke%
}{%
David%
\ \protect \BOthers {.}}{%
{\protect \APACyear {2001}}%
}]{%
990595}
\APACinsertmetastar {%
990595}%
\begin{APACrefauthors}%
David, L.%
, Cottet, F.%
\BCBL {}\ \BBA {} Nissanke, N.%
\end{APACrefauthors}%
\unskip\
\newblock
\APACrefYearMonthDay{2001}{}{}.
\newblock
{\BBOQ}\APACrefatitle {Jitter control in on-line scheduling of dependent
  real-time tasks} {Jitter control in on-line scheduling of dependent real-time
  tasks}.{\BBCQ}
\newblock
\BIn{} \APACrefbtitle {Proceedings 22nd IEEE Real-Time Systems Symposium (RTSS
  2001) (Cat. No.01PR1420)} {Proceedings 22nd ieee real-time systems symposium
  (rtss 2001) (cat. no.01pr1420)}\ (\BPG~49-58).
\newblock
\begin{APACrefDOI} \doi{10.1109/REAL.2001.990595} \end{APACrefDOI}
\PrintBackRefs{\CurrentBib}

\bibitem [\protect \citeauthoryear {%
Dawson%
, Wellman%
\BCBL {}\ \BBA {} Anderson%
}{%
Dawson%
\ \protect \BOthers {.}}{%
{\protect \APACyear {2010}}%
}]{%
dawson2010}
\APACinsertmetastar {%
dawson2010}%
\begin{APACrefauthors}%
Dawson, S.%
, Wellman, B\BPBI L.%
\BCBL {}\ \BBA {} Anderson, M.%
\end{APACrefauthors}%
\unskip\
\newblock
\APACrefYearMonthDay{2010}{}{}.
\newblock
{\BBOQ}\APACrefatitle {Using simulation to predict multi-robot performance on
  coverage tasks} {Using simulation to predict multi-robot performance on
  coverage tasks}.{\BBCQ}
\newblock
\BIn{} \APACrefbtitle {2010 IEEE/RSJ International Conference on Intelligent
  Robots and Systems} {2010 ieee/rsj international conference on intelligent
  robots and systems}\ (\BPGS\ 202--208).
\newblock
\begin{APACrefDOI} \doi{10.1109/IROS.2010.5650679} \end{APACrefDOI}
\PrintBackRefs{\CurrentBib}

\bibitem [\protect \citeauthoryear {%
Dear%
\ \BBA {} Sherif%
}{%
Dear%
\ \BBA {} Sherif%
}{%
{\protect \APACyear {2000}}%
}]{%
dear2000}
\APACinsertmetastar {%
dear2000}%
\begin{APACrefauthors}%
Dear, R\BPBI G.%
\BCBT {}\ \BBA {} Sherif, J\BPBI S.%
\end{APACrefauthors}%
\unskip\
\newblock
\APACrefYearMonthDay{2000}{}{}.
\newblock
{\BBOQ}\APACrefatitle {Using Simulation To Evaluate Resource Utilization
  Strategies} {Using simulation to evaluate resource utilization
  strategies}.{\BBCQ}
\newblock
\APACjournalVolNumPages{SIMULATION}{74}{2}{75--83}.
\newblock
\begin{APACrefURL} \url{https://doi.org/10.1177/003754970007400202}
  \end{APACrefURL}
\newblock
\begin{APACrefDOI} \doi{10.1177/003754970007400202} \end{APACrefDOI}
\PrintBackRefs{\CurrentBib}

\bibitem [\protect \citeauthoryear {%
Dorri%
, Kanhere%
\BCBL {}\ \BBA {} Jurdak%
}{%
Dorri%
\ \protect \BOthers {.}}{%
{\protect \APACyear {2018}}%
}]{%
8352646}
\APACinsertmetastar {%
8352646}%
\begin{APACrefauthors}%
Dorri, A.%
, Kanhere, S\BPBI S.%
\BCBL {}\ \BBA {} Jurdak, R.%
\end{APACrefauthors}%
\unskip\
\newblock
\APACrefYearMonthDay{2018}{}{}.
\newblock
{\BBOQ}\APACrefatitle {Multi-Agent Systems: A Survey} {Multi-agent systems: A
  survey}.{\BBCQ}
\newblock
\APACjournalVolNumPages{IEEE Access}{6}{}{28573-28593}.
\newblock
\begin{APACrefDOI} \doi{10.1109/ACCESS.2018.2831228} \end{APACrefDOI}
\PrintBackRefs{\CurrentBib}

\bibitem [\protect \citeauthoryear {%
{Eramo}%
\ \protect \BOthers {.}}{%
{Eramo}%
\ \protect \BOthers {.}}{%
{\protect \APACyear {2016}}%
}]{%
Eramo2016}
\APACinsertmetastar {%
Eramo2016}%
\begin{APACrefauthors}%
{Eramo}, V.%
, Listanti, M.%
, Lavacca, F\BPBI G.%
, Iovanna, P.%
, Bottari, G.%
\BCBL {}\ \BBA {} Ponzini, F.%
\end{APACrefauthors}%
\unskip\
\newblock
\APACrefYearMonthDay{2016}{}{}.
\newblock
{\BBOQ}\APACrefatitle {Trade-Off Between Power and Bandwidth Consumption in a
  Reconfigurable Xhaul Network Architecture} {Trade-off between power and
  bandwidth consumption in a reconfigurable xhaul network architecture}.{\BBCQ}
\newblock
\APACjournalVolNumPages{IEEE Access}{4}{}{9053--9065}.
\newblock
\begin{APACrefDOI} \doi{10.1109/ACCESS.2016.2639578} \end{APACrefDOI}
\PrintBackRefs{\CurrentBib}

\bibitem [\protect \citeauthoryear {%
Fiscko%
, Kar%
\BCBL {}\ \BBA {} Sinopoli%
}{%
Fiscko%
\ \protect \BOthers {.}}{%
{\protect \APACyear {2021}}%
}]{%
9482994}
\APACinsertmetastar {%
9482994}%
\begin{APACrefauthors}%
Fiscko, C.%
, Kar, S.%
\BCBL {}\ \BBA {} Sinopoli, B.%
\end{APACrefauthors}%
\unskip\
\newblock
\APACrefYearMonthDay{2021}{}{}.
\newblock
{\BBOQ}\APACrefatitle {Efficient Solutions for Targeted Control of Multi-Agent
  MDPs} {Efficient solutions for targeted control of multi-agent mdps}.{\BBCQ}
\newblock
\BIn{} \APACrefbtitle {2021 American Control Conference (ACC)} {2021 american
  control conference (acc)}\ (\BPG~690-696).
\newblock
\begin{APACrefDOI} \doi{10.23919/ACC50511.2021.9482994} \end{APACrefDOI}
\PrintBackRefs{\CurrentBib}

\bibitem [\protect \citeauthoryear {%
Ghassemi%
, DePauw%
\BCBL {}\ \BBA {} Chowdhury%
}{%
Ghassemi%
\ \protect \BOthers {.}}{%
{\protect \APACyear {2019}}%
}]{%
8901062}
\APACinsertmetastar {%
8901062}%
\begin{APACrefauthors}%
Ghassemi, P.%
, DePauw, D.%
\BCBL {}\ \BBA {} Chowdhury, S.%
\end{APACrefauthors}%
\unskip\
\newblock
\APACrefYearMonthDay{2019}{}{}.
\newblock
{\BBOQ}\APACrefatitle {Decentralized Dynamic Task Allocation in Swarm Robotic
  Systems for Disaster Response: Extended Abstract} {Decentralized dynamic task
  allocation in swarm robotic systems for disaster response: Extended
  abstract}.{\BBCQ}
\newblock
\BIn{} \APACrefbtitle {2019 International Symposium on Multi-Robot and
  Multi-Agent Systems (MRS)} {2019 international symposium on multi-robot and
  multi-agent systems (mrs)}\ (\BPG~83-85).
\newblock
\begin{APACrefDOI} \doi{10.1109/MRS.2019.8901062} \end{APACrefDOI}
\PrintBackRefs{\CurrentBib}

\bibitem [\protect \citeauthoryear {%
Gruler%
, Fikar%
, Juan%
, Hirsch%
\BCBL {}\ \BBA {} Contreras-Bolton%
}{%
Gruler%
\ \protect \BOthers {.}}{%
{\protect \APACyear {2017}}%
}]{%
gruler2017supporting}
\APACinsertmetastar {%
gruler2017supporting}%
\begin{APACrefauthors}%
Gruler, A.%
, Fikar, C.%
, Juan, A\BPBI A.%
, Hirsch, P.%
\BCBL {}\ \BBA {} Contreras-Bolton, C.%
\end{APACrefauthors}%
\unskip\
\newblock
\APACrefYearMonthDay{2017}{}{}.
\newblock
{\BBOQ}\APACrefatitle {Supporting multi-depot and stochastic waste collection
  management in clustered urban areas via simulation--optimization} {Supporting
  multi-depot and stochastic waste collection management in clustered urban
  areas via simulation--optimization}.{\BBCQ}
\newblock
\APACjournalVolNumPages{Journal of simulation}{11}{1}{11--19}.
\PrintBackRefs{\CurrentBib}

\bibitem [\protect \citeauthoryear {%
{Guo}%
\ \BBA {} {Dimarogonas}%
}{%
{Guo}%
\ \BBA {} {Dimarogonas}%
}{%
{\protect \APACyear {2017}}%
}]{%
Guo2017}
\APACinsertmetastar {%
Guo2017}%
\begin{APACrefauthors}%
{Guo}, M.%
\BCBT {}\ \BBA {} {Dimarogonas}, D\BPBI V.%
\end{APACrefauthors}%
\unskip\
\newblock
\APACrefYearMonthDay{2017}{}{}.
\newblock
{\BBOQ}\APACrefatitle {Task and Motion Coordination for Heterogeneous
  Multiagent Systems With Loosely Coupled Local Tasks} {Task and motion
  coordination for heterogeneous multiagent systems with loosely coupled local
  tasks}.{\BBCQ}
\newblock
\APACjournalVolNumPages{IEEE Transactions on Automation Science and
  Engineering}{14}{2}{797-808}.
\newblock
\begin{APACrefDOI} \doi{10.1109/TASE.2016.2628389} \end{APACrefDOI}
\PrintBackRefs{\CurrentBib}

\bibitem [\protect \citeauthoryear {%
{Gupta}%
\ \BBA {} {Pujari}%
}{%
{Gupta}%
\ \BBA {} {Pujari}%
}{%
{\protect \APACyear {2009}}%
}]{%
health2009}
\APACinsertmetastar {%
health2009}%
\begin{APACrefauthors}%
{Gupta}, S.%
\BCBT {}\ \BBA {} {Pujari}, S.%
\end{APACrefauthors}%
\unskip\
\newblock
\APACrefYearMonthDay{2009}{}{}.
\newblock
{\BBOQ}\APACrefatitle {A multi-agent system (MAS) based scheme for health care
  and medical diagnosis system} {A multi-agent system (mas) based scheme for
  health care and medical diagnosis system}.{\BBCQ}
\newblock
\BIn{} \APACrefbtitle {c International Conference on Intelligent Agent
  Multi-Agent Systems} {c international conference on intelligent agent
  multi-agent systems}\ (\BPG~1-3).
\newblock
\begin{APACrefDOI} \doi{10.1109/IAMA.2009.5228086} \end{APACrefDOI}
\PrintBackRefs{\CurrentBib}

\bibitem [\protect \citeauthoryear {%
Hill%
\ \BBA {} Marty%
}{%
Hill%
\ \BBA {} Marty%
}{%
{\protect \APACyear {2008}}%
}]{%
Hill2008}
\APACinsertmetastar {%
Hill2008}%
\begin{APACrefauthors}%
Hill, M\BPBI D.%
\BCBT {}\ \BBA {} Marty, M\BPBI R.%
\end{APACrefauthors}%
\unskip\
\newblock
\APACrefYearMonthDay{2008}{}{}.
\newblock
{\BBOQ}\APACrefatitle {{Amdahl's Law in the Multicore Era}} {{Amdahl's Law in
  the Multicore Era}}.{\BBCQ}
\newblock
\APACjournalVolNumPages{Computer}{41}{7}{33--38}.
\newblock
\begin{APACrefDOI} \doi{10.1109/MC.2008.209} \end{APACrefDOI}
\PrintBackRefs{\CurrentBib}

\bibitem [\protect \citeauthoryear {%
{Hongwei An}%
, {Xiong Li}%
\BCBL {}\ \BBA {} {Xiuquan Xie}%
}{%
{Hongwei An}%
\ \protect \BOthers {.}}{%
{\protect \APACyear {2010}}%
}]{%
battlefield2010}
\APACinsertmetastar {%
battlefield2010}%
\begin{APACrefauthors}%
{Hongwei An}%
, {Xiong Li}%
\BCBL {}\ \BBA {} {Xiuquan Xie}.%
\end{APACrefauthors}%
\unskip\
\newblock
\APACrefYearMonthDay{2010}{}{}.
\newblock
{\BBOQ}\APACrefatitle {Multi-agent interactions centric virtual battlefield
  simulation model} {Multi-agent interactions centric virtual battlefield
  simulation model}.{\BBCQ}
\newblock
\BIn{} \APACrefbtitle {2010 2nd International Conference on Advanced Computer
  Control} {2010 2nd international conference on advanced computer control}\
  (\BVOL~3, \BPG~315-319).
\newblock
\begin{APACrefDOI} \doi{10.1109/ICACC.2010.5486850} \end{APACrefDOI}
\PrintBackRefs{\CurrentBib}

\bibitem [\protect \citeauthoryear {%
Hubmann%
, Schulz%
, Becker%
, Althoff%
\BCBL {}\ \BBA {} Stiller%
}{%
Hubmann%
\ \protect \BOthers {.}}{%
{\protect \APACyear {2018}}%
}]{%
8248668}
\APACinsertmetastar {%
8248668}%
\begin{APACrefauthors}%
Hubmann, C.%
, Schulz, J.%
, Becker, M.%
, Althoff, D.%
\BCBL {}\ \BBA {} Stiller, C.%
\end{APACrefauthors}%
\unskip\
\newblock
\APACrefYearMonthDay{2018}{}{}.
\newblock
{\BBOQ}\APACrefatitle {Automated Driving in Uncertain Environments: Planning
  With Interaction and Uncertain Maneuver Prediction} {Automated driving in
  uncertain environments: Planning with interaction and uncertain maneuver
  prediction}.{\BBCQ}
\newblock
\APACjournalVolNumPages{IEEE Transactions on Intelligent Vehicles}{3}{1}{5-17}.
\newblock
\begin{APACrefDOI} \doi{10.1109/TIV.2017.2788208} \end{APACrefDOI}
\PrintBackRefs{\CurrentBib}

\bibitem [\protect \citeauthoryear {%
{Ismail}%
, {Shaikh Ali}%
\BCBL {}\ \BBA {} {Abu Bakar}%
}{%
{Ismail}%
\ \protect \BOthers {.}}{%
{\protect \APACyear {2018}}%
}]{%
Selfregulated2018}
\APACinsertmetastar {%
Selfregulated2018}%
\begin{APACrefauthors}%
{Ismail}, S.%
, {Shaikh Ali}, S\BPBI H.%
\BCBL {}\ \BBA {} {Abu Bakar}, M\BPBI H.%
\end{APACrefauthors}%
\unskip\
\newblock
\APACrefYearMonthDay{2018}{}{}.
\newblock
{\BBOQ}\APACrefatitle {Agent-based Self-regulated Learning Simulation Adopting
  the Concept of GUSC Model} {Agent-based self-regulated learning simulation
  adopting the concept of gusc model}.{\BBCQ}
\newblock
\BIn{} \APACrefbtitle {2018 International Symposium on Agent, Multi-Agent
  Systems and Robotics (ISAMSR)} {2018 international symposium on agent,
  multi-agent systems and robotics (isamsr)}\ (\BPG~1-6).
\newblock
\begin{APACrefDOI} \doi{10.1109/ISAMSR.2018.8540556} \end{APACrefDOI}
\PrintBackRefs{\CurrentBib}

\bibitem [\protect \citeauthoryear {%
Kurniawati%
, Hsu%
\BCBL {}\ \BBA {} Lee%
}{%
Kurniawati%
\ \protect \BOthers {.}}{%
{\protect \APACyear {2008}}%
}]{%
kurniawati2008sarsop}
\APACinsertmetastar {%
kurniawati2008sarsop}%
\begin{APACrefauthors}%
Kurniawati, H.%
, Hsu, D.%
\BCBL {}\ \BBA {} Lee, W\BPBI S.%
\end{APACrefauthors}%
\unskip\
\newblock
\APACrefYearMonthDay{2008}{}{}.
\newblock
{\BBOQ}\APACrefatitle {Sarsop: Efficient point-based pomdp planning by
  approximating optimally reachable belief spaces.} {Sarsop: Efficient
  point-based pomdp planning by approximating optimally reachable belief
  spaces.}{\BBCQ}
\newblock
\BIn{} \APACrefbtitle {Robotics: Science and systems} {Robotics: Science and
  systems}\ (\BVOL\ 2008).
\PrintBackRefs{\CurrentBib}

\bibitem [\protect \citeauthoryear {%
S.~Lee%
, Jain%
\BCBL {}\ \BBA {} Son%
}{%
S.~Lee%
\ \protect \BOthers {.}}{%
{\protect \APACyear {2022}}%
}]{%
lee2022}
\APACinsertmetastar {%
lee2022}%
\begin{APACrefauthors}%
Lee, S.%
, Jain, S.%
\BCBL {}\ \BBA {} Son, Y\BHBI J.%
\end{APACrefauthors}%
\unskip\
\newblock
\APACrefYearMonthDay{2022}{{\APACmonth{01}}}{}.
\newblock
{\BBOQ}\APACrefatitle {A Hierarchical Decision-Making Framework in Social
  Networks for Efficient Disaster Management} {A hierarchical decision-making
  framework in social networks for efficient disaster management}.{\BBCQ}
\newblock
\APACjournalVolNumPages{ACM Trans. Model. Comput. Simul.}{32}{1}{}.
\newblock
\begin{APACrefURL} \url{https://doi.org/10.1145/3490027} \end{APACrefURL}
\newblock
\begin{APACrefDOI} \doi{10.1145/3490027} \end{APACrefDOI}
\PrintBackRefs{\CurrentBib}

\bibitem [\protect \citeauthoryear {%
Y\BPBI C.~Lee%
\ \BBA {} Zomaya%
}{%
Y\BPBI C.~Lee%
\ \BBA {} Zomaya%
}{%
{\protect \APACyear {2011}}%
}]{%
Lee2011}
\APACinsertmetastar {%
Lee2011}%
\begin{APACrefauthors}%
Lee, Y\BPBI C.%
\BCBT {}\ \BBA {} Zomaya, A\BPBI Y.%
\end{APACrefauthors}%
\unskip\
\newblock
\APACrefYearMonthDay{2011}{{\APACmonth{08}}}{}.
\newblock
{\BBOQ}\APACrefatitle {Energy Conscious Scheduling for Distributed Computing
  Systems under Different Operating Conditions} {Energy conscious scheduling
  for distributed computing systems under different operating
  conditions}.{\BBCQ}
\newblock
\APACjournalVolNumPages{IEEE Transactions on Parallel and Distributed
  Systems}{22}{8}{1374--1381}.
\newblock
\APACrefnote{doi:10.1109/TPDS.2010.208}
\newblock
\begin{APACrefDOI} \doi{10.1109/TPDS.2010.208} \end{APACrefDOI}
\PrintBackRefs{\CurrentBib}

\bibitem [\protect \citeauthoryear {%
{Li}%
, {Maddah-Ali}%
, {Yu}%
\BCBL {}\ \BBA {} {Avestimehr}%
}{%
{Li}%
\ \protect \BOthers {.}}{%
{\protect \APACyear {2018}}%
}]{%
Li2018}
\APACinsertmetastar {%
Li2018}%
\begin{APACrefauthors}%
{Li}, S.%
, {Maddah-Ali}, M\BPBI A.%
, {Yu}, Q.%
\BCBL {}\ \BBA {} {Avestimehr}, A\BPBI S.%
\end{APACrefauthors}%
\unskip\
\newblock
\APACrefYearMonthDay{2018}{{\APACmonth{01}}}{}.
\newblock
{\BBOQ}\APACrefatitle {A Fundamental Tradeoff Between Computation and
  Communication in Distributed Computing} {A fundamental tradeoff between
  computation and communication in distributed computing}.{\BBCQ}
\newblock
\APACjournalVolNumPages{IEEE Transactions on Information
  Theory}{64}{1}{109--128}.
\newblock
\begin{APACrefDOI} \doi{10.1109/TIT.2017.2756959} \end{APACrefDOI}
\PrintBackRefs{\CurrentBib}

\bibitem [\protect \citeauthoryear {%
Littman%
, Dean%
\BCBL {}\ \BBA {} Kaelbling%
}{%
Littman%
\ \protect \BOthers {.}}{%
{\protect \APACyear {2013}}%
}]{%
littman2013complexity}
\APACinsertmetastar {%
littman2013complexity}%
\begin{APACrefauthors}%
Littman, M\BPBI L.%
, Dean, T\BPBI L.%
\BCBL {}\ \BBA {} Kaelbling, L\BPBI P.%
\end{APACrefauthors}%
\unskip\
\newblock
\APACrefYearMonthDay{2013}{}{}.
\newblock
{\BBOQ}\APACrefatitle {On the complexity of solving Markov decision problems}
  {On the complexity of solving markov decision problems}.{\BBCQ}
\newblock
\APACjournalVolNumPages{arXiv preprint arXiv:1302.4971}{}{}{}.
\PrintBackRefs{\CurrentBib}

\bibitem [\protect \citeauthoryear {%
C.~Liu%
\ \protect \BOthers {.}}{%
C.~Liu%
\ \protect \BOthers {.}}{%
{\protect \APACyear {2016}}%
}]{%
Liu2016}
\APACinsertmetastar {%
Liu2016}%
\begin{APACrefauthors}%
Liu, C.%
, Hamrick, J\BPBI B.%
, Fisac, J\BPBI F.%
, Dragan, A\BPBI D.%
, Hedrick, J\BPBI K.%
, Sastry, S\BPBI S.%
\BCBL {}\ \BBA {} Griffiths, T\BPBI L.%
\end{APACrefauthors}%
\unskip\
\newblock
\APACrefYearMonthDay{2016}{}{}.
\newblock
{\BBOQ}\APACrefatitle {{Goal Inference Improves Objective and Perceived
  Performance in Human-Robot Collaboration}} {{Goal Inference Improves
  Objective and Perceived Performance in Human-Robot Collaboration}}.{\BBCQ}
\newblock
\BIn{} C\BPBI M.~Jonker, S.~Marsella, J.~Thangarajah\BCBL {}\ \BBA {} K.~Tuyls\
  (\BEDS), \APACrefbtitle {Proceedings of the 2016 International Conference on
  Autonomous Agents {\&} Multiagent Systems, Singapore, May 9-13, 2016}
  {Proceedings of the 2016 international conference on autonomous agents {\&}
  multiagent systems, singapore, may 9-13, 2016}\ (\BPGS\ 940--948).
\newblock
\APACaddressPublisher{}{{ACM}}.
\PrintBackRefs{\CurrentBib}

\bibitem [\protect \citeauthoryear {%
F.~Liu%
\ \BBA {} Liu%
}{%
F.~Liu%
\ \BBA {} Liu%
}{%
{\protect \APACyear {2018}}%
}]{%
8576124}
\APACinsertmetastar {%
8576124}%
\begin{APACrefauthors}%
Liu, F.%
\BCBT {}\ \BBA {} Liu, Z.%
\end{APACrefauthors}%
\unskip\
\newblock
\APACrefYearMonthDay{2018}{}{}.
\newblock
{\BBOQ}\APACrefatitle {A Neighborhood-Based Value Iteration Algorithm for POMDP
  Problems} {A neighborhood-based value iteration algorithm for pomdp
  problems}.{\BBCQ}
\newblock
\BIn{} \APACrefbtitle {2018 IEEE 30th International Conference on Tools with
  Artificial Intelligence (ICTAI)} {2018 ieee 30th international conference on
  tools with artificial intelligence (ictai)}\ (\BPG~808-812).
\newblock
\begin{APACrefDOI} \doi{10.1109/ICTAI.2018.00126} \end{APACrefDOI}
\PrintBackRefs{\CurrentBib}

\bibitem [\protect \citeauthoryear {%
{Liu}%
\ \protect \BOthers {.}}{%
{Liu}%
\ \protect \BOthers {.}}{%
{\protect \APACyear {2020}}%
}]{%
Coordination2020}
\APACinsertmetastar {%
Coordination2020}%
\begin{APACrefauthors}%
{Liu}, M.%
, {Chang}, W.%
, {Li}, C.%
, {Ji}, Y.%
, {Li}, R.%
\BCBL {}\ \BBA {} {Feng}, M.%
\end{APACrefauthors}%
\unskip\
\newblock
\APACrefYearMonthDay{2020}{}{}.
\newblock
{\BBOQ}\APACrefatitle {Discrete Interactions in Decentralized Multiagent
  Coordination: A Probabilistic Perspective} {Discrete interactions in
  decentralized multiagent coordination: A probabilistic perspective}.{\BBCQ}
\newblock
\APACjournalVolNumPages{IEEE Transactions on Cognitive and Developmental
  Systems}{}{}{1-1}.
\newblock
\begin{APACrefDOI} \doi{10.1109/TCDS.2020.3040769} \end{APACrefDOI}
\PrintBackRefs{\CurrentBib}

\bibitem [\protect \citeauthoryear {%
Lu%
, Nolte%
, Bate%
\BCBL {}\ \BBA {} Norström%
}{%
Lu%
, Nolte%
, Bate%
\BCBL {}\ \BBA {} Norström%
}{%
{\protect \APACyear {2010}}%
}]{%
5676303}
\APACinsertmetastar {%
5676303}%
\begin{APACrefauthors}%
Lu, Y.%
, Nolte, T.%
, Bate, I.%
\BCBL {}\ \BBA {} Norström, C.%
\end{APACrefauthors}%
\unskip\
\newblock
\APACrefYearMonthDay{2010}{}{}.
\newblock
{\BBOQ}\APACrefatitle {Timing Analyzing for Systems with Task Execution
  Dependencies} {Timing analyzing for systems with task execution
  dependencies}.{\BBCQ}
\newblock
\BIn{} \APACrefbtitle {2010 IEEE 34th Annual Computer Software and Applications
  Conference} {2010 ieee 34th annual computer software and applications
  conference}\ (\BPG~515-524).
\newblock
\begin{APACrefDOI} \doi{10.1109/COMPSAC.2010.57} \end{APACrefDOI}
\PrintBackRefs{\CurrentBib}

\bibitem [\protect \citeauthoryear {%
Lu%
, Nolte%
, Kraft%
\BCBL {}\ \BBA {} Norstrom%
}{%
Lu%
, Nolte%
, Kraft%
\BCBL {}\ \BBA {} Norstrom%
}{%
{\protect \APACyear {2010}}%
}]{%
5628617}
\APACinsertmetastar {%
5628617}%
\begin{APACrefauthors}%
Lu, Y.%
, Nolte, T.%
, Kraft, J.%
\BCBL {}\ \BBA {} Norstrom, C.%
\end{APACrefauthors}%
\unskip\
\newblock
\APACrefYearMonthDay{2010}{}{}.
\newblock
{\BBOQ}\APACrefatitle {Statistical-Based Response-Time Analysis of Systems with
  Execution Dependencies between Tasks} {Statistical-based response-time
  analysis of systems with execution dependencies between tasks}.{\BBCQ}
\newblock
\BIn{} \APACrefbtitle {2010 15th IEEE International Conference on Engineering
  of Complex Computer Systems} {2010 15th ieee international conference on
  engineering of complex computer systems}\ (\BPG~169-179).
\newblock
\begin{APACrefDOI} \doi{10.1109/ICECCS.2010.55} \end{APACrefDOI}
\PrintBackRefs{\CurrentBib}

\bibitem [\protect \citeauthoryear {%
Mahela%
\ \protect \BOthers {.}}{%
Mahela%
\ \protect \BOthers {.}}{%
{\protect \APACyear {2020}}%
}]{%
9299490}
\APACinsertmetastar {%
9299490}%
\begin{APACrefauthors}%
Mahela, O\BPBI P.%
, Khosravy, M.%
, Gupta, N.%
, Khan, B.%
, Alhelou, H\BPBI H.%
, Mahla, R.%
\BDBL {}Siano, P.%
\end{APACrefauthors}%
\unskip\
\newblock
\APACrefYearMonthDay{2020}{}{}.
\newblock
{\BBOQ}\APACrefatitle {Comprehensive overview of multi-agent systems for
  controlling smart grids} {Comprehensive overview of multi-agent systems for
  controlling smart grids}.{\BBCQ}
\newblock
\APACjournalVolNumPages{CSEE Journal of Power and Energy Systems}{}{}{1-16}.
\newblock
\begin{APACrefDOI} \doi{10.17775/CSEEJPES.2020.03390} \end{APACrefDOI}
\PrintBackRefs{\CurrentBib}

\bibitem [\protect \citeauthoryear {%
Merton%
}{%
Merton%
}{%
{\protect \APACyear {1968}}%
}]{%
merton1968}
\APACinsertmetastar {%
merton1968}%
\begin{APACrefauthors}%
Merton, R\BPBI K.%
\end{APACrefauthors}%
\unskip\
\newblock
\APACrefYearMonthDay{1968}{}{}.
\newblock
{\BBOQ}\APACrefatitle {The Matthew effect in science: The reward and
  communication systems of science are considered} {The matthew effect in
  science: The reward and communication systems of science are
  considered}.{\BBCQ}
\newblock
\APACjournalVolNumPages{Science}{159}{3810}{56--63}.
\PrintBackRefs{\CurrentBib}

\bibitem [\protect \citeauthoryear {%
Mukhopadhyay%
\ \BBA {} Jain%
}{%
Mukhopadhyay%
\ \BBA {} Jain%
}{%
{\protect \APACyear {2001}}%
}]{%
971476}
\APACinsertmetastar {%
971476}%
\begin{APACrefauthors}%
Mukhopadhyay, S.%
\BCBT {}\ \BBA {} Jain, B.%
\end{APACrefauthors}%
\unskip\
\newblock
\APACrefYearMonthDay{2001}{}{}.
\newblock
{\BBOQ}\APACrefatitle {Multi-agent Markov decision processes with limited agent
  communication} {Multi-agent markov decision processes with limited agent
  communication}.{\BBCQ}
\newblock
\BIn{} \APACrefbtitle {Proceeding of the 2001 IEEE International Symposium on
  Intelligent Control (ISIC '01) (Cat. No.01CH37206)} {Proceeding of the 2001
  ieee international symposium on intelligent control (isic '01) (cat.
  no.01ch37206)}\ (\BPG~7-12).
\newblock
\begin{APACrefDOI} \doi{10.1109/ISIC.2001.971476} \end{APACrefDOI}
\PrintBackRefs{\CurrentBib}

\bibitem [\protect \citeauthoryear {%
Ndoye%
\ \BBA {} Sorel%
}{%
Ndoye%
\ \BBA {} Sorel%
}{%
{\protect \APACyear {2013}}%
}]{%
6755290}
\APACinsertmetastar {%
6755290}%
\begin{APACrefauthors}%
Ndoye, F.%
\BCBT {}\ \BBA {} Sorel, Y.%
\end{APACrefauthors}%
\unskip\
\newblock
\APACrefYearMonthDay{2013}{}{}.
\newblock
{\BBOQ}\APACrefatitle {Monoprocessor Real-Time Scheduling of Data Dependent
  Tasks with Exact Preemption Cost for Embedded Systems} {Monoprocessor
  real-time scheduling of data dependent tasks with exact preemption cost for
  embedded systems}.{\BBCQ}
\newblock
\BIn{} \APACrefbtitle {2013 IEEE 16th International Conference on Computational
  Science and Engineering} {2013 ieee 16th international conference on
  computational science and engineering}\ (\BPG~714-721).
\newblock
\begin{APACrefDOI} \doi{10.1109/CSE.2013.110} \end{APACrefDOI}
\PrintBackRefs{\CurrentBib}

\bibitem [\protect \citeauthoryear {%
{Qin}%
, {Ouyang}%
\BCBL {}\ \BBA {} {Xiong}%
}{%
{Qin}%
\ \protect \BOthers {.}}{%
{\protect \APACyear {2018}}%
}]{%
Qin2018}
\APACinsertmetastar {%
Qin2018}%
\begin{APACrefauthors}%
{Qin}, L.%
, {Ouyang}, F.%
\BCBL {}\ \BBA {} {Xiong}, G.%
\end{APACrefauthors}%
\unskip\
\newblock
\APACrefYearMonthDay{2018}{}{}.
\newblock
{\BBOQ}\APACrefatitle {Dependent task scheduling algorithm in distributed
  system} {Dependent task scheduling algorithm in distributed system}.{\BBCQ}
\newblock
\BIn{} \APACrefbtitle {2018 4th International Conference on Computer and
  Technology Applications (ICCTA)} {2018 4th international conference on
  computer and technology applications (iccta)}\ (\BPG~91-95).
\newblock
\begin{APACrefDOI} \doi{10.1109/CATA.2018.8398662} \end{APACrefDOI}
\PrintBackRefs{\CurrentBib}

\bibitem [\protect \citeauthoryear {%
Rigney%
}{%
Rigney%
}{%
{\protect \APACyear {2010}}%
}]{%
rigney2010}
\APACinsertmetastar {%
rigney2010}%
\begin{APACrefauthors}%
Rigney, D.%
\end{APACrefauthors}%
\unskip\
\newblock
\APACrefYear{2010}.
\newblock
\APACrefbtitle {The Matthew effect: How advantage begets further advantage}
  {The matthew effect: How advantage begets further advantage}.
\newblock
\APACaddressPublisher{}{Columbia University Press}.
\PrintBackRefs{\CurrentBib}

\bibitem [\protect \citeauthoryear {%
Salmer{\'o}n-Garc{\i}%
, Inigo-Blasco%
, D{\i}%
, Cagigas-Muniz%
\BCBL {}\ \protect \BOthers {.}}{%
Salmer{\'o}n-Garc{\i}%
\ \protect \BOthers {.}}{%
{\protect \APACyear {2015}}%
}]{%
Garcia2015}
\APACinsertmetastar {%
Garcia2015}%
\begin{APACrefauthors}%
Salmer{\'o}n-Garc{\i}, J.%
, Inigo-Blasco, P.%
, D{\i}, F.%
, Cagigas-Muniz, D.%
\BCBL {}\ \BOthersPeriod {.}\end{APACrefauthors}%
\unskip\
\newblock
\APACrefYearMonthDay{2015}{}{}.
\newblock
{\BBOQ}\APACrefatitle {A tradeoff analysis of a cloud-based robot navigation
  assistant using stereo image processing} {A tradeoff analysis of a
  cloud-based robot navigation assistant using stereo image processing}.{\BBCQ}
\newblock
\APACjournalVolNumPages{IEEE Transactions on Automation Science and
  Engineering}{12}{2}{444--454}.
\PrintBackRefs{\CurrentBib}

\bibitem [\protect \citeauthoryear {%
Shani%
, Pineau%
\BCBL {}\ \BBA {} Kaplow%
}{%
Shani%
\ \protect \BOthers {.}}{%
{\protect \APACyear {2013}}%
}]{%
shani2013survey}
\APACinsertmetastar {%
shani2013survey}%
\begin{APACrefauthors}%
Shani, G.%
, Pineau, J.%
\BCBL {}\ \BBA {} Kaplow, R.%
\end{APACrefauthors}%
\unskip\
\newblock
\APACrefYearMonthDay{2013}{}{}.
\newblock
{\BBOQ}\APACrefatitle {A survey of point-based POMDP solvers} {A survey of
  point-based pomdp solvers}.{\BBCQ}
\newblock
\APACjournalVolNumPages{Autonomous Agents and Multi-Agent
  Systems}{27}{1}{1--51}.
\PrintBackRefs{\CurrentBib}

\bibitem [\protect \citeauthoryear {%
Shi%
, Ueter%
, von~der Brüggen%
\BCBL {}\ \BBA {} Chen%
}{%
Shi%
\ \protect \BOthers {.}}{%
{\protect \APACyear {2019}}%
}]{%
8743172}
\APACinsertmetastar {%
8743172}%
\begin{APACrefauthors}%
Shi, J.%
, Ueter, N.%
, von~der Brüggen, G.%
\BCBL {}\ \BBA {} Chen, J\BHBI j.%
\end{APACrefauthors}%
\unskip\
\newblock
\APACrefYearMonthDay{2019}{}{}.
\newblock
{\BBOQ}\APACrefatitle {Multiprocessor Synchronization of Periodic Real-Time
  Tasks Using Dependency Graphs} {Multiprocessor synchronization of periodic
  real-time tasks using dependency graphs}.{\BBCQ}
\newblock
\BIn{} \APACrefbtitle {2019 IEEE Real-Time and Embedded Technology and
  Applications Symposium (RTAS)} {2019 ieee real-time and embedded technology
  and applications symposium (rtas)}\ (\BPG~279-292).
\newblock
\begin{APACrefDOI} \doi{10.1109/RTAS.2019.00031} \end{APACrefDOI}
\PrintBackRefs{\CurrentBib}

\bibitem [\protect \citeauthoryear {%
Squazzoni%
\ \BBA {} Gandelli%
}{%
Squazzoni%
\ \BBA {} Gandelli%
}{%
{\protect \APACyear {2012}}%
}]{%
squazzoni2011}
\APACinsertmetastar {%
squazzoni2011}%
\begin{APACrefauthors}%
Squazzoni, F.%
\BCBT {}\ \BBA {} Gandelli, C.%
\end{APACrefauthors}%
\unskip\
\newblock
\APACrefYearMonthDay{2012}{}{}.
\newblock
{\BBOQ}\APACrefatitle {{Saint Matthew strikes again: An agent-based model of
  peer review and the scientific community structure}} {{Saint Matthew strikes
  again: An agent-based model of peer review and the scientific community
  structure}}.{\BBCQ}
\newblock
\APACjournalVolNumPages{Journal of Informetrics}{6}{2}{265-275}.
\newblock
\begin{APACrefURL}
  \url{https://ideas.repec.org/a/eee/infome/v6y2012i2p265-275.html}
  \end{APACrefURL}
\newblock
\begin{APACrefDOI} \doi{10.1016/j.joi.2011.12.005} \end{APACrefDOI}
\PrintBackRefs{\CurrentBib}

\bibitem [\protect \citeauthoryear {%
Tavanpour%
, Kazi%
\BCBL {}\ \BBA {} Wainer%
}{%
Tavanpour%
\ \protect \BOthers {.}}{%
{\protect \APACyear {2020}}%
}]{%
tavanpour2020discrete}
\APACinsertmetastar {%
tavanpour2020discrete}%
\begin{APACrefauthors}%
Tavanpour, M.%
, Kazi, B\BPBI U.%
\BCBL {}\ \BBA {} Wainer, G.%
\end{APACrefauthors}%
\unskip\
\newblock
\APACrefYearMonthDay{2020}{}{}.
\newblock
{\BBOQ}\APACrefatitle {Discrete Event Systems Specifications Modelling and
  Simulation of Wireless Networking Applications} {Discrete event systems
  specifications modelling and simulation of wireless networking
  applications}.{\BBCQ}
\newblock
\APACjournalVolNumPages{Journal of Simulation}{}{}{1--25}.
\PrintBackRefs{\CurrentBib}

\bibitem [\protect \citeauthoryear {%
{Topcuoglu}%
, {Hariri}%
\BCBL {}\ \BBA {} {Min-You Wu}%
}{%
{Topcuoglu}%
\ \protect \BOthers {.}}{%
{\protect \APACyear {2002}}%
}]{%
Topcuoglu2002}
\APACinsertmetastar {%
Topcuoglu2002}%
\begin{APACrefauthors}%
{Topcuoglu}, H.%
, {Hariri}, S.%
\BCBL {}\ \BBA {} {Min-You Wu}.%
\end{APACrefauthors}%
\unskip\
\newblock
\APACrefYearMonthDay{2002}{}{}.
\newblock
{\BBOQ}\APACrefatitle {Performance-effective and low-complexity task scheduling
  for heterogeneous computing} {Performance-effective and low-complexity task
  scheduling for heterogeneous computing}.{\BBCQ}
\newblock
\APACjournalVolNumPages{IEEE Transactions on Parallel and Distributed
  Systems}{13}{3}{260-274}.
\newblock
\begin{APACrefDOI} \doi{10.1109/71.993206} \end{APACrefDOI}
\PrintBackRefs{\CurrentBib}

\bibitem [\protect \citeauthoryear {%
Viseras%
, Xu%
\BCBL {}\ \BBA {} Merino%
}{%
Viseras%
\ \protect \BOthers {.}}{%
{\protect \APACyear {2020}}%
}]{%
viseras2020}
\APACinsertmetastar {%
viseras2020}%
\begin{APACrefauthors}%
Viseras, A.%
, Xu, Z.%
\BCBL {}\ \BBA {} Merino, L.%
\end{APACrefauthors}%
\unskip\
\newblock
\APACrefYearMonthDay{2020}{}{}.
\newblock
{\BBOQ}\APACrefatitle {Distributed Multi-Robot Information Gathering under
  Spatio-Temporal Inter-Robot Constraints} {Distributed multi-robot information
  gathering under spatio-temporal inter-robot constraints}.{\BBCQ}
\newblock
\APACjournalVolNumPages{Sensors}{20}{2}{}.
\newblock
\begin{APACrefURL} \url{https://www.mdpi.com/1424-8220/20/2/484}
  \end{APACrefURL}
\newblock
\begin{APACrefDOI} \doi{10.3390/s20020484} \end{APACrefDOI}
\PrintBackRefs{\CurrentBib}

\bibitem [\protect \citeauthoryear {%
Vlassis%
, Spaan%
\BCBL {}\ \protect \BOthers {.}}{%
Vlassis%
\ \protect \BOthers {.}}{%
{\protect \APACyear {2004}}%
}]{%
vlassis2004fast}
\APACinsertmetastar {%
vlassis2004fast}%
\begin{APACrefauthors}%
Vlassis, N.%
, Spaan, M\BPBI T.%
\BCBL {}\ \BOthersPeriod {.}\end{APACrefauthors}%
\unskip\
\newblock
\APACrefYearMonthDay{2004}{}{}.
\newblock
{\BBOQ}\APACrefatitle {A fast point-based algorithm for POMDPs} {A fast
  point-based algorithm for pomdps}.{\BBCQ}
\newblock
\BIn{} \APACrefbtitle {Benelearn 2004: Proceedings of the Annual Machine
  Learning Conference of Belgium and the Netherlands} {Benelearn 2004:
  Proceedings of the annual machine learning conference of belgium and the
  netherlands}\ (\BPGS\ 170--176).
\PrintBackRefs{\CurrentBib}

\bibitem [\protect \citeauthoryear {%
Waeber%
, Frazier%
\BCBL {}\ \BBA {} Henderson%
}{%
Waeber%
\ \protect \BOthers {.}}{%
{\protect \APACyear {2012}}%
}]{%
waeber2012}
\APACinsertmetastar {%
waeber2012}%
\begin{APACrefauthors}%
Waeber, R.%
, Frazier, P\BPBI I.%
\BCBL {}\ \BBA {} Henderson, S\BPBI G.%
\end{APACrefauthors}%
\unskip\
\newblock
\APACrefYearMonthDay{2012}{aug}{}.
\newblock
{\BBOQ}\APACrefatitle {A Framework for Selecting a Selection Procedure} {A
  framework for selecting a selection procedure}.{\BBCQ}
\newblock
\APACjournalVolNumPages{ACM Trans. Model. Comput. Simul.}{22}{3}{}.
\newblock
\begin{APACrefURL} \url{https://doi.org/10.1145/2331140.2331144}
  \end{APACrefURL}
\newblock
\begin{APACrefDOI} \doi{10.1145/2331140.2331144} \end{APACrefDOI}
\PrintBackRefs{\CurrentBib}

\bibitem [\protect \citeauthoryear {%
Wang%
, Zhou%
, Li%
\BCBL {}\ \BBA {} Shan%
}{%
Wang%
\ \protect \BOthers {.}}{%
{\protect \APACyear {2018}}%
}]{%
wang2018impact}
\APACinsertmetastar {%
wang2018impact}%
\begin{APACrefauthors}%
Wang, J.%
, Zhou, W.%
, Li, S.%
\BCBL {}\ \BBA {} Shan, D.%
\end{APACrefauthors}%
\unskip\
\newblock
\APACrefYearMonthDay{2018}{}{}.
\newblock
{\BBOQ}\APACrefatitle {Impact of personalised route recommendation in the
  cooperation vehicle-infrastructure systems on the network traffic flow
  evolution} {Impact of personalised route recommendation in the cooperation
  vehicle-infrastructure systems on the network traffic flow evolution}.{\BBCQ}
\newblock
\APACjournalVolNumPages{Journal of Simulation}{}{}{}.
\PrintBackRefs{\CurrentBib}

\bibitem [\protect \citeauthoryear {%
Wilsdorf%
, Pierce%
, Hillston%
\BCBL {}\ \BBA {} Uhrmacher%
}{%
Wilsdorf%
\ \protect \BOthers {.}}{%
{\protect \APACyear {2019}}%
}]{%
wilsdorf2019}
\APACinsertmetastar {%
wilsdorf2019}%
\begin{APACrefauthors}%
Wilsdorf, P.%
, Pierce, M\BPBI E.%
, Hillston, J.%
\BCBL {}\ \BBA {} Uhrmacher, A\BPBI M.%
\end{APACrefauthors}%
\unskip\
\newblock
\APACrefYearMonthDay{2019}{}{}.
\newblock
{\BBOQ}\APACrefatitle {Round-Based Super-Individuals---Balancing Speed and
  Accuracy} {Round-based super-individuals---balancing speed and
  accuracy}.{\BBCQ}
\newblock
\BIn{} \APACrefbtitle {Proceedings of the 2019 ACM SIGSIM Conference on
  Principles of Advanced Discrete Simulation} {Proceedings of the 2019 acm
  sigsim conference on principles of advanced discrete simulation}\
  (\BPG~95?98).
\newblock
\APACaddressPublisher{New York, NY, USA}{Association for Computing Machinery}.
\newblock
\begin{APACrefURL} \url{https://doi.org/10.1145/3316480.3322894}
  \end{APACrefURL}
\newblock
\begin{APACrefDOI} \doi{10.1145/3316480.3322894} \end{APACrefDOI}
\PrintBackRefs{\CurrentBib}

\bibitem [\protect \citeauthoryear {%
Xiao%
\ \BBA {} Peng%
}{%
Xiao%
\ \BBA {} Peng%
}{%
{\protect \APACyear {2019}}%
}]{%
Xiao2019}
\APACinsertmetastar {%
Xiao2019}%
\begin{APACrefauthors}%
Xiao, J.%
\BCBT {}\ \BBA {} Peng, J.%
\end{APACrefauthors}%
\unskip\
\newblock
\APACrefYearMonthDay{2019}{{\APACmonth{07}}}{}.
\newblock
{\BBOQ}\APACrefatitle {Trade-offs between computation, communication, and
  synchronization in stencil-collective alternate update} {Trade-offs between
  computation, communication, and synchronization in stencil-collective
  alternate update}.{\BBCQ}
\newblock
\APACjournalVolNumPages{CCF Transactions on High Performance
  Computing}{}{1}{144--160}.
\newblock
\begin{APACrefDOI} \doi{10.1007/s42514-019-00011-x} \end{APACrefDOI}
\PrintBackRefs{\CurrentBib}

\bibitem [\protect \citeauthoryear {%
{Xu}%
\ \BBA {} {Yang}%
}{%
{Xu}%
\ \BBA {} {Yang}%
}{%
{\protect \APACyear {2009}}%
}]{%
Dynamic2009}
\APACinsertmetastar {%
Dynamic2009}%
\begin{APACrefauthors}%
{Xu}, H.%
\BCBT {}\ \BBA {} {Yang}, Y.%
\end{APACrefauthors}%
\unskip\
\newblock
\APACrefYearMonthDay{2009}{}{}.
\newblock
{\BBOQ}\APACrefatitle {Research and Design on Dynamic Multi-agent Cooperative
  Processing Model} {Research and design on dynamic multi-agent cooperative
  processing model}.{\BBCQ}
\newblock
\BIn{} \APACrefbtitle {2009 International Conference on Web Information Systems
  and Mining} {2009 international conference on web information systems and
  mining}\ (\BPG~432-436).
\newblock
\begin{APACrefDOI} \doi{10.1109/WISM.2009.94} \end{APACrefDOI}
\PrintBackRefs{\CurrentBib}

\bibitem [\protect \citeauthoryear {%
{Yang}%
, {Yu}%
, {Liu}%
, {Wang}%
\BCBL {}\ \BBA {} {Guo}%
}{%
{Yang}%
\ \protect \BOthers {.}}{%
{\protect \APACyear {2019}}%
}]{%
Crowdsourcing2019}
\APACinsertmetastar {%
Crowdsourcing2019}%
\begin{APACrefauthors}%
{Yang}, C.%
, {Yu}, Z.%
, {Liu}, Y.%
, {Wang}, L.%
\BCBL {}\ \BBA {} {Guo}, B.%
\end{APACrefauthors}%
\unskip\
\newblock
\APACrefYearMonthDay{2019}{}{}.
\newblock
{\BBOQ}\APACrefatitle {Dynamic Allocation for Complex Mobile Crowdsourcing Task
  with Internal Dependencies} {Dynamic allocation for complex mobile
  crowdsourcing task with internal dependencies}.{\BBCQ}
\newblock
\BIn{} \APACrefbtitle {2019 IEEE SmartWorld, Ubiquitous Intelligence Computing,
  Advanced Trusted Computing, Scalable Computing Communications, Cloud Big Data
  Computing, Internet of People and Smart City Innovation
  (SmartWorld/SCALCOM/UIC/ATC/CBDCom/IOP/SCI)} {2019 ieee smartworld,
  ubiquitous intelligence computing, advanced trusted computing, scalable
  computing communications, cloud big data computing, internet of people and
  smart city innovation (smartworld/scalcom/uic/atc/cbdcom/iop/sci)}\
  (\BPG~818-825).
\newblock
\begin{APACrefDOI} \doi{10.1109/SmartWorld-UIC-ATC-SCALCOM-IOP-SCI.2019.00171}
  \end{APACrefDOI}
\PrintBackRefs{\CurrentBib}

\bibitem [\protect \citeauthoryear {%
J.~Zhang%
, Wei%
, Liu%
\BCBL {}\ \BBA {} Deng%
}{%
J.~Zhang%
\ \protect \BOthers {.}}{%
{\protect \APACyear {2021}}%
}]{%
zhang2021}
\APACinsertmetastar {%
zhang2021}%
\begin{APACrefauthors}%
Zhang, J.%
, Wei, L.%
, Liu, M.%
\BCBL {}\ \BBA {} Deng, Y.%
\end{APACrefauthors}%
\unskip\
\newblock
\APACrefYearMonthDay{2021}{}{}.
\newblock
{\BBOQ}\APACrefatitle {A Competition Model for Modeling and Describing Matthew
  Effect in Computational Social Systems} {A competition model for modeling and
  describing matthew effect in computational social systems}.{\BBCQ}
\newblock
\BIn{} \APACrefbtitle {2021 11th International Conference on Intelligent
  Control and Information Processing (ICICIP)} {2021 11th international
  conference on intelligent control and information processing (icicip)}\
  (\BPG~438-443).
\newblock
\begin{APACrefDOI} \doi{10.1109/ICICIP53388.2021.9642207} \end{APACrefDOI}
\PrintBackRefs{\CurrentBib}

\bibitem [\protect \citeauthoryear {%
Z.~Zhang%
, Hsu%
\BCBL {}\ \BBA {} Lee%
}{%
Z.~Zhang%
\ \protect \BOthers {.}}{%
{\protect \APACyear {2014}}%
}]{%
zhang2014covering}
\APACinsertmetastar {%
zhang2014covering}%
\begin{APACrefauthors}%
Zhang, Z.%
, Hsu, D.%
\BCBL {}\ \BBA {} Lee, W\BPBI S.%
\end{APACrefauthors}%
\unskip\
\newblock
\APACrefYearMonthDay{2014}{}{}.
\newblock
{\BBOQ}\APACrefatitle {Covering number for efficient heuristic-based POMDP
  planning} {Covering number for efficient heuristic-based pomdp
  planning}.{\BBCQ}
\newblock
\BIn{} \APACrefbtitle {International conference on machine learning}
  {International conference on machine learning}\ (\BPGS\ 28--36).
\PrintBackRefs{\CurrentBib}

\bibitem [\protect \citeauthoryear {%
L.~{Zhao}%
, {Du}%
\BCBL {}\ \BBA {} {Chen}%
}{%
L.~{Zhao}%
\ \protect \BOthers {.}}{%
{\protect \APACyear {2018}}%
}]{%
Zhao2018}
\APACinsertmetastar {%
Zhao2018}%
\begin{APACrefauthors}%
{Zhao}, L.%
, {Du}, M.%
\BCBL {}\ \BBA {} {Chen}, L.%
\end{APACrefauthors}%
\unskip\
\newblock
\APACrefYearMonthDay{2018}{}{}.
\newblock
{\BBOQ}\APACrefatitle {A new multi-resource allocation mechanism: A tradeoff
  between fairness and efficiency in cloud computing} {A new multi-resource
  allocation mechanism: A tradeoff between fairness and efficiency in cloud
  computing}.{\BBCQ}
\newblock
\APACjournalVolNumPages{China Communications}{15}{3}{57-77}.
\newblock
\begin{APACrefDOI} \doi{10.1109/CC.2018.8331991} \end{APACrefDOI}
\PrintBackRefs{\CurrentBib}

\bibitem [\protect \citeauthoryear {%
Y.~{Zhao}%
, {Chen}%
, {Li}%
\BCBL {}\ \BBA {} {Tian}%
}{%
Y.~{Zhao}%
\ \protect \BOthers {.}}{%
{\protect \APACyear {2014}}%
}]{%
Zhao2014}
\APACinsertmetastar {%
Zhao2014}%
\begin{APACrefauthors}%
{Zhao}, Y.%
, {Chen}, L.%
, {Li}, Y.%
\BCBL {}\ \BBA {} {Tian}, W.%
\end{APACrefauthors}%
\unskip\
\newblock
\APACrefYearMonthDay{2014}{}{}.
\newblock
{\BBOQ}\APACrefatitle {Efficient task scheduling for Many Task Computing with
  resource attribute selection} {Efficient task scheduling for many task
  computing with resource attribute selection}.{\BBCQ}
\newblock
\APACjournalVolNumPages{China Communications}{11}{12}{125-140}.
\newblock
\begin{APACrefDOI} \doi{10.1109/CC.2014.7019847} \end{APACrefDOI}
\PrintBackRefs{\CurrentBib}

\bibitem [\protect \citeauthoryear {%
{Zomaya}%
\ \BBA {} {Lee}%
}{%
{Zomaya}%
\ \BBA {} {Lee}%
}{%
{\protect \APACyear {2012}}%
}]{%
Zomaya2012}
\APACinsertmetastar {%
Zomaya2012}%
\begin{APACrefauthors}%
{Zomaya}, A\BPBI Y.%
\BCBT {}\ \BBA {} {Lee}, Y\BPBI C.%
\end{APACrefauthors}%
\unskip\
\newblock
\APACrefYearMonthDay{2012}{}{}.
\newblock
{\BBOQ}\APACrefatitle {Comparison and Analysis of Greedy Energy-Efficient
  Scheduling Algorithms for Computational Grids} {Comparison and analysis of
  greedy energy-efficient scheduling algorithms for computational
  grids}.{\BBCQ}
\newblock
\BIn{} \APACrefbtitle {Energy-Efficient Distributed Computing Systems}
  {Energy-efficient distributed computing systems}\ (\BPG~189-214).
\newblock
\begin{APACrefDOI} \doi{10.1002/9781118342015.ch7} \end{APACrefDOI}
\PrintBackRefs{\CurrentBib}

\end{thebibliography}
\end{document}